\documentclass[12pt,a4paper]{article}

\usepackage{amsmath}
\usepackage{amssymb}
\usepackage[nosort]{cite}
\usepackage[hyperref,bulletsep]{collect}
\def\gfxon{\usepackage[final]{graphicx}}

\gfxon

\makeatletter
\let\old@makecaption=\@makecaption
\def\@makecaption{\small\old@makecaption}
\makeatother

\makeatletter
\@addtoreset{equation}{section}
\makeatother

\makeatletter
\let\old@startsection=\@startsection
\renewcommand{\@startsection}[6]{\old@startsection{#1}{#2}{#3}{#4}{#5}{#6\mathversion{bold}}}
\makeatother

\let\oldPhi=\Phi
\let\oldPsi=\Psi
\let\oldGamma=\Gamma
\let\oldSigma=\Sigma
\let\oldDelta=\Delta
\let\oldPi=\Pi
\renewcommand{\Phi}{\mathnormal{\oldPhi}}
\renewcommand{\Psi}{\mathnormal{\oldPsi}}
\renewcommand{\Gamma}{\mathnormal{\oldGamma}}
\renewcommand{\Sigma}{\mathnormal{\oldSigma}}
\renewcommand{\Delta}{\mathnormal{\oldDelta}}
\renewcommand{\Pi}{\mathnormal{\oldPi}}

\newcommand{\figref}[1]{Fig.~\ref{#1}}
\newcommand{\secref}[1]{Sec.~\ref{#1}}

\newcommand{\tabref}[1]{Tab.~\ref{#1}}

\newcommand{\sfrac}[2]{{\textstyle\frac{#1}{#2}}}
\newcommand{\half}{\sfrac{1}{2}}

\newcommand{\alg}[1]{\mathfrak{#1}}
\newcommand{\grp}[1]{\mathrm{#1}}
\newcommand{\grSU}{\grp{SU}}

\newcommand{\grU}{\grp{U}}
\newcommand{\grSO}{\grp{SO}}
\newcommand{\grSp}{\grp{Sp}}
\newcommand{\grPSU}{\grp{PSU}}
\newcommand{\alU}{\alg{u}}
\newcommand{\alSU}{\alg{su}}

\newcommand{\alSL}{\alg{sl}}
\newcommand{\alSO}{\alg{so}}

\newcommand{\alPSU}{\alg{psu}}

\newcommand{\order}[1]{\mathcal{O}(#1)}

\newcommand{\superN}{\mathcal{N}}
\newcommand{\gym}{g_{\scriptscriptstyle\mathrm{YM}}}

\newcommand{\Tr}{\mathop{\mathrm{Tr}}}
\newcommand{\eng}{\mathop{\mathrm{eng}}}
\newcommand{\rep}[1]{{\mathbf{#1}}}
\newcommand{\indup}[1]{_{\mathrm{#1}}}

\newcommand{\Real}{\mathbb{R}}
\newcommand{\Integers}{\mathbb{Z}}
\newcommand{\matr}[2]{\left(\begin{array}{#1}#2\end{array}\right)}
\newcommand{\Op}{\mathcal{O}}
\newcommand{\charge}{\mathcal{Q}}
\newcommand{\transfer}{\mathcal{T}}

\newcommand{\state}[1]{|#1\rangle}%
\newcommand{\bigstate}[1]{\big|#1\big\rangle}%
\newcommand{\PTerm}[2]{\big\{{\textstyle\genfrac{}{}{0pt}{}{#1}{#2}}\big\}}

\newcommand{\lrbrk}[1]{\left(#1\right)}
\newcommand{\bigbrk}[1]{\bigl(#1\bigr)}

\newcommand{\bigcomm}[2]{\big[#1,#2\big]}
\newcommand{\comm}[2]{[#1,#2]}

\newcommand{\acomm}[2]{\{#1,#2\}}

\newcommand{\bigeval}[1]{#1\big|}

\newcommand{\nln}{\nonumber\\}
\newcommand{\nl}{\nonumber\\&&\mathord{}}
\newcommand{\nlnum}{\\&&\mathord{}}
\newcommand{\earel}[1]{\mathrel{}&#1&\mathrel{}}
\newcommand{\eq}{\earel{=}}
\newenvironment{myeqnarray}{\arraycolsep0pt\begin{eqnarray}}{\end{eqnarray}\ignorespacesafterend}
\newenvironment{myeqnarray*}{\arraycolsep0pt\begin{eqnarray*}}{\end{eqnarray*}\ignorespacesafterend}

\def\[{\begin{equation}}
\def\]{\end{equation}}
\def\<{\begin{myeqnarray}}
\def\>{\end{myeqnarray}}

\ifx\href\asklfhas\newcommand{\href}[2]{#2}\fi
\newcommand{\arxivno}[1]{\href{http://arxiv.org/abs/#1}{#1}}

\begin{document}

%\begin{comment}

\setcounter{page}{0}
\thispagestyle{empty}
\begin{flushright}\footnotesize
\texttt{\arxivno{hep-th/0310252}}\\
\texttt{AEI 2003-087}
\end{flushright}
\vspace{2cm}

\begin{center}
{\Large\textbf{\mathversion{bold}
The $\alSU(2|3)$ Dynamic Spin Chain
}\par}
\vspace{2cm}

\textsc{Niklas Beisert}
\vspace{5mm}

\textit{Max-Planck-Institut f\"ur Gravitationsphysik\\
Albert-Einstein-Institut\\
Am M\"uhlenberg 1, 14476 Potsdam, Germany}
\vspace{3mm}

\texttt{nbeisert@aei.mpg.de}\par\vspace*{2cm}\vspace*{\fill}

\textbf{Abstract}\vspace{7mm}

\begin{minipage}{12.7cm}\small
The complete one-loop, planar dilatation operator of the
$\superN=4$ superconformal gauge theory was recently 
derived and shown to be integrable. 
Here, we present further compelling evidence for 
a generalisation of this integrable structure to higher orders
of the coupling constant. 
For that we consider the $\alSU(2|3)$ subsector and investigate 
the restrictions imposed on the spin chain Hamiltonian by 
the symmetry algebra. This allows us to uniquely
fix the energy shifts up to the three-loop level
and thus prove the correctness of a conjecture in 
\arxivno{hep-th/0303060}.
A novel aspect of this spin chain model is that 
the higher-loop Hamiltonian,
as for $\superN=4$ SYM in general, does not preserve the 
number of spin sites. Yet this 
\emph{dynamic} spin chain appears to be integrable. \par
\end{minipage}\vspace*{\fill}

\end{center}

\newpage
\setcounter{footnote}{0}

%\end{comment}

%%%%%%%%%%%%%%%%%%%%%%%%%%%%%%%%%%%%%%%%%%%%%%%%%%%%%%%%%%%%%%%%%%%%%%%%%%%%%%%%
\section{Introduction and conclusions}
\label{sec:Intro}

%\remark{figures}
%
A couple of years ago, string theory and the AdS/CFT correspondence 
\cite{Maldacena:1998re,Witten:1998qj,Gubser:1998bc}
have renewed the interest in $\superN=4$ Super Yang-Mills theory 
\cite{Gliozzi:1977qd,Brink:1977bc}. 
In particular, the BMN correspondence \cite{Berenstein:2002jq}
has attracted attention to scaling dimensions of local operators
and efficient methods for their computation have been developed
\cite{Kristjansen:2002bb,Constable:2002hw,Beisert:2002bb,Constable:2002vq,Beisert:2002ff}.
In this context Minahan and Zarembo found that 
the planar one-loop dilatation operator in the scalar sector of $\superN=4$ SYM 
is isomorphic to the Hamiltonian of an integrable $\alSO(6)$ spin chain \cite{Minahan:2002ve}.
Integrable spin chains with $\alSL(2)$ symmetry have been discovered in 
(non-supersymmetric) gauge theories before
\cite{Lipatov:1994yb,Faddeev:1995zg,Braun:1998id,Belitsky:1999qh,Braun:1999te,Belitsky:1999ru,Belitsky:1999bf,Derkachov:1999ze}.
The main difference between these two developments is
that the symmetry algebras correspond to internal 
and spacetime symmetries, respectively.
In a supersymmetric theory the internal and spacetime symmetries join
and, for $\superN=4$ SYM, form the supergroup $\grPSU(2,2|4)$.
Quite naturally the complete one-loop planar dilatation generator of this theory
\cite{Beisert:2003jj} corresponds to an integrable super spin chain 
with full $\alPSU(2,2|4)$ symmetry \cite{Beisert:2003yb}
(for literature on super spin chains, c.f.~\cite{Saleur:1999cx} and references therein).
For further aspects of integrability in $\superN=4$ SYM 
such as a discovery of a Yangian enhancement of the superconformal algebra 
and relations to Wilson-loops, the cusp anomaly and gravity,
c.f.~\cite{Belitsky:2003ys,Dolan:2003uh,Gorsky:2003nq}.
%For further aspects of integrability in $\superN=4$ SYM 
%(Yangian, Wilson-loops) see
%\cite{Belitsky:2003ys,Dolan:2003uh,Gorsky:2003nq}.
\medskip

For an integrable spin chain there exists an $R$-matrix 
satisfying the Yang-Baxter equation.
From this one can derive an infinite number of 
charges $\charge_n$ which mutually commute
\[\comm{\charge_m}{\charge_n}=0.\]
The first non-trivial of these is the Hamiltonian $H=\charge_2$. 
One, and possibly the only, directly observable effect 
of this symmetry enhancement is the existence 
of \emph{parity pairs} \cite{Beisert:2003tq}. 
The energies $E_\pm$ of certain pairs of states 
with opposite parity eigenvalues coincide, 
\[
E_+= E_-,
\]
a fact which cannot be attributed to any of the obvious symmetries. 
In this context `parity' refers to a
$\Integers_2$ symmetry of $\grSU(N)$ gauge theory.

This curiosity is merely the tip of an iceberg,
for integrability opens the gates for very precise tests 
of the AdS/CFT correspondence.
It is no longer necessary to compute and diagonalise 
the matrix of anomalous dimensions, 
a very non-trivial task for long spin chains.
Instead, one may use the Bethe ansatz 
(c.f.~\cite{Faddeev:1996iy} for a pedagogical introduction)
to obtain the one-loop anomalous dimensions directly \cite{Minahan:2002ve,Beisert:2003yb}.
In the thermodynamic limit, which would be practically inaccessible by conventional methods,
the algebraic Bethe equations turn into integral equations. 
With the Bethe ansatz at hand, it became possible to compute anomalous dimensions
of operators with large spin quantum numbers
\cite{Beisert:2003xu,Beisert:2003ea,Engquist:2003rn,Arutyunov:2003rg}.
Via the AdS/CFT correspondence these states correspond to 
highly spinning string configurations. 
Even though quantisation of string theory on $AdS_5\times S^5$ is an open problem,
these spinning strings can be treated in a semiclassical fashion.
String theory results
\cite{Frolov:2002av,Russo:2002sr,Minahan:2002rc,Tseytlin:2002ny,Frolov:2003qc,Frolov:2003tu,Frolov:2003xy,Arutyunov:2003uj}, 
largely due to Frolov and Tseytlin,
(see also \cite{Gubser:2002tv,Alishahiha:2002fi,Alishahiha:2003xe,Mateos:2003de,Schvellinger:2003vz})
were shown to agree perfectly with gauge theory in a very
intricate way \cite{Beisert:2003xu,Beisert:2003ea,Engquist:2003rn,Arutyunov:2003rg},
and thus provide novel, compelling evidence 
in favour of the AdS/CFT correspondence.
\bigskip

Now, one might object that this would be valid at \emph{one-loop} only.
The Hamiltonian of an integrable spin chain is usually of 
\emph{nearest-neighbour} type (as for one-loop gauge theories) or, 
at least, involves only two, non-neighbouring spins at a time. 
In contrast, higher order corrections to the dilatation generator
require interactions of \emph{more than two fields}.
Moreover, the \emph{number of fields is not even conserved} in general. 
To the author's knowledge, spin chains with any of these features 
have largely been neglected up to now%
\footnote{The higher charges of an integrable spin chain
are indeed of this type. Nevertheless, they cannot yield 
the higher-loop corrections, because they commute among themselves,
whereas the higher-loop corrections do not.}.
Yet, it would be fascinating if they existed. 
In \cite{Beisert:2003tq} the two-loop dilatation generator 
was derived in a subsector of $\superN=4$ SYM, 
the $\alSU(2)$ subsector.
Indeed, it was observed that parity pairs
were preserved, i.e.~that their \emph{two-loop} anomalous 
dimensions remain \emph{degenerate}
\[
\oldDelta_+(\lambda)=\oldDelta_-(\lambda).
\]
This `miracle' was subsequently explained by a 
generalisation of the higher charges $\charge_{n}$ 
to higher orders $\charge_n(\lambda)$ 
of the 't~Hooft coupling constant $\lambda=\gym^2N$.
For a few of the higher charges it turned out to be possible 
to do this in such a way as to preserve
their abelian algebra
\[
\comm{\charge_m(\lambda)}{\charge_n(\lambda)}=0
\]
in a perturbative sense, i.e.~up to higher order terms in 
the coupling constant. 
Short of a suitable $R$-matrix and Yang-Baxter equation
for the system, this relation remains the best guess for a 
definition of higher-loop integrability.
What is more, the analysis of \cite{Beisert:2003tq}
showed that integrability can be maintained to, at least, 
the four-loop level and suggested that, 
as long as no degeneracies are broken, 
the complete tower of one-loop commuting charges can be 
extended to higher loops.
\medskip

However, due to the simplicity of the
$\alSU(2)$ subsector, there are two limitations to this observation.
Firstly, any admissible two-loop contribution turns out to be integrable
and one should be careful in drawing conclusions about higher-loops from this.
Secondly, the number of spin sites is conserved. 
Even if one should find higher-loop integrability in the $\alSU(2)$ 
subsector, this might turn out not to generalise to the full theory.
On the other hand, there is one clear indication for integrability
by means of the AdS/CFT correspondence. 
For instance, the appearance of a flat current in the 
$AdS_5\times S^5$ string sigma model gives rise to a Yangian structure
\cite{Mandal:2002fs,Bena:2003wd,Vallilo:2003nx,Alday:2003zb}
which is related to integrability. 
The analysis of semiclassical strings showed that indeed one 
can find an infinite tower of commuting charges \cite{Arutyunov:2003rg}
and indeed they agree, at one-loop, with the spin chain charges~\cite{Arutyunov:2003rg}. 
Interestingly, the string theory charges
are analytic functions of the effective string tension.
What could these quantities correspond to if 
$\superN=4$ SYM were not integrable?
%The analysis of semiclassical strings showed that indeed one 
%can find an infinite tower of commuting charges 
%$\mathcal{Q}_k(\lambda)$ \cite{Arutyunov:2003rg}
%and indeed they agree with the spin chain charges $Q_k$ at
%the one-loop level \cite{Arutyunov:2003rg}. 
%Interestingly, the charges $\mathcal{Q}_k(\lambda)$ obtained in string theory are 
%valid to all orders of the 't~Hooft coupling constant. 
%What could these all-loop quantities correspond to if not to $Q_k(\lambda)$,
%i.e.~if $\superN=4$ SYM were not integrable at all values of the coupling constant?

Although higher-loop integrability is an extremely interesting 
prospect, obviously, it is also very non-trivial to confirm or disprove
in a direct way.
In \cite{Beisert:2003tq,Beisert:2003jb} we have made an educated
guess for the dilatation generator at three-loops and four-loops. 
This conjecture crucially depended on 
the assumption of higher-loop integrability.
Unfortunately, a calculation in near plane wave string theory \cite{Callan:2003xr}
produces a result which is apparently in contradiction 
with the three-loop gauge theory conjecture. 
Thus, it may seem that integrability would have to break down at three-loops.
\bigskip

The aim of the current work is to derive the dilatation operator of $\superN=4$ SYM
at three-loops and thus shed further light on the issue of higher-loop integrability. 
As there is no reasonable chance for a straight field theoretic computation
we will have to rely on other firm facts. Here, we will use
the symmetry algebra and structural limitations from field theory
to constrain the result. 
We will not make the \emph{assumption of higher-loop integrability},
in fact it will be the \emph{outcome}.
To finally match our result with ${\superN=4}$ SYM
we will demand correct BMN scaling behaviour \cite{Berenstein:2002jq,Gross:2002su,Santambrogio:2002sb}.
Consequently, we are able to \emph{prove} the correctness of the 
conjectured three-loop dilatation operator of \cite{Beisert:2003tq}%
\footnote{We do not have a cure to the disagreement noticed in \cite{Callan:2003xr}.}. 
%and pass the buck back.}.
In conclusion, a direct computation does not only appear hopeless,
it is rather unnecessary. 

In this paper we restrict to the $\alSU(2|3)$ closed subsector of
$\superN=4$ SYM \cite{Beisert:2003jj}%
\footnote{This is also a closed subsector of the BMN matrix 
model \cite{Berenstein:2002jq}. 
The results apply equally well to a 
perturbative analysis as in \cite{Kim:2002if,Kim:2003rz,Klose:2003qc}.
Indeed the three-loop result for the $\alSU(2)$ subsector is compatible 
with the result of Klose and Plefka \cite{Klose:2003qc}. This implies that 
not only the $\alSU(2)$ subsector, but rather the complete 
$\alSU(2|3)$ subsector of the BMN matrix model coincides 
with $\superN=4$ SYM up to a redefinition of the coupling constant.}.
In comparison to the $\alSU(2)$ subsector 
(which is contained in this model) there are two key improvements.
Firstly, the dilatation generator is part of the symmetry (super)algebra,
the closure of which imposes more serious constraints on the available structures. 
This is what allows us to go to higher loop orders and rely on less assumptions.
Secondly, the number of spin sites is allowed to fluctuate,
the spin chain becomes \emph{dynamic}. 
Both of these novel features are present in $\superN=4$ gauge theory 
%(the dynamic aspect is related to the Konishi anomaly \cite{Konishi:1984hf,Konishi:1985tu})
and make the $\alSU(2|3)$ subsector a more realistic model. 
We hope that this investigation will help us to understand the higher-loop 
corrections in $\superN=4$ SYM and we feel
that a generalisation of the results to the full theory should be possible.
Apart from the gauge theory point of view, the model in itself is very exciting
due to its unexplored characteristics.
\bigskip

The model under consideration is a spin chain 
with spins in the fundamental representation $\rep{3|2}$
of $\alSU(2|3)$.
The free symmetry generators $J^A_0$ acting on the spin chain 
Hilbert space are constructed as tensor product
representations of $\rep{3|2}$'s.
The strategy will be to consider deformations $J^A(g)$ of the
algebra generators, $g\sim \sqrt{\lambda}=\gym\sqrt{N}$, such that the
algebra relations remain unchanged
\[\label{eq:Comm}
\comm{J^A(g)}{J^B(g)}=f^{AB}_C J^C(g).
\]
Note that in contrast to quantum algebras, the quantum corrections change
only the representation $J^A(g)$. The algebra remains classical.
Together with structural results from field theory we
are able to fix the Hamiltonian $H(g)$ up to three-loops, 
i.e.~sixth order in $g$. In this model, not only the Hamiltonian
receives corrections, but also the supercharges $Q(g),S(g)$,
and we can fix their deformations up to fourth order.%
\medskip

This paper is organised as follows:
We start in \secref{sec:Model} by describing the model. 
In \secref{sec:OneLoop} we proceed by 
considering the Hamiltonian up to one-loop order
and thus illustrate our procedure in a simple context.
We will then extend the results to two-loops and three-loops
in \secref{sec:TwoLoop},~\ref{sec:ThreeLoop}.
In \secref{sec:Spec} we apply 
the Hamiltonian to the first few states in the
spectrum and comment on integrability.
Finally, we summarise some open questions in \secref{sec:Outlook}.

%%%%%%%%%%%%%%%%%%%%%%%%%%%%%%%%%%%%%%%%%%%%%%%%%%%%%%%%%%%%%%%%%%%%%%%%%%%%%%%%
\section{The Model}
\label{sec:Model}

In this section we describe the model in terms of the 
space of states, symmetry and how it is related to $\superN=4$ 
gauge theory.

%%%%%%%%%%%%%%%%%%%%%%%%%%%%%%%%%%%%%%%%%%%%%%%%%%%%%%%%%%%%%%%%%%%%%%%%%%%%%
\subsection{A Subsector of $\superN=4$ SYM}

In this paper we consider single-trace local operators in
a subsector $\superN=4$ SYM. 
In a gauge theory local operators are composed as
gauge invariant combinations of the
fundamental fields (and derivatives thereof). 
For a gauge group $\grU(N)$ the fields are $N\times N$ matrices and 
gauge invariant objects are constructed by taking traces
of products of these matrices. 
Generically, there are infinitely many fields 
(due to the derivatives) to be chosen from.

\paragraph{Fields.}
To simplify the investigation 
we will restrict to a finite subset of fields. 
In particular we choose the $\alSU(2|3)$ subsector.
As was shown in \cite{Beisert:2003jj}, 
the truncation to this subsector does not interfere 
with the mixing patterns of local operators and is therefore consistent
to all orders of perturbation theory. It also represents the maximal closed subsector
with finitely many fields.
The subsector consists of three complex scalars $\phi_a$ 
(Latin indices take the values $1,2,3$)
and two complex fermions $\psi_\alpha$ 
(Greek indices take the values $1,2$)%
\footnote{Equivalently, we might start out with the BMN matrix model
\cite{Berenstein:2002jq} and restrict to its $\alSU(2|3)$ subsector.}
\[
\phi_a\quad (a=1,2,3),\qquad
\psi_\alpha\quad (\alpha=1,2).
\]
These can be combined into a supermultiplet $W_A$ 
(capital indices range from $1$ to $5$)
of fields 
\[W_{1,2,3}=\phi_{1,2,3},\quad
W_{4,5}=\psi_{1,2}.\]
\paragraph{States.}
A single-trace operator is a linear combination of basis states
\[
\label{eq:TraceState}
\state{A_1\ldots A_L}=\Tr W_{A_1}\ldots W_{A_L}.\]
Note that due to cyclicity of the trace we have to identify states 
as follows
\[\label{eq:TraceCycl}
\state{A_1\ldots A_L}=(-1)^{(A_1\ldots A_k)(A_{k+1}\ldots A_L)}\state{A_{k+1}\ldots A_L A_1\ldots A_k},
\]
where $(-1)^{XY}$ equals $-1$ if both $X$ and $Y$ are fermionic; $+1$ otherwise.
In particular \eqref{eq:TraceCycl} implies that states which can be written as
\[\label{eq:TraceFerm}
\bigstate{(A_1\ldots A_{L/2})^{2}}=-\bigstate{(A_1\ldots A_{L/2})^{2}}=0
\]
do not exist if $A_1\ldots A_{L/2}$ contains an odd number of fermions.
A generic local operator of this subsector is a linear combination
of the basis states \eqref{eq:TraceState}.

\paragraph{Parity.}
States can be classified by their parity.
Parity inverts the order of fields within the trace according to
\[\label{eq:TraceParity}
P\state{A_1\ldots A_L}=(-1)^{L+k(k+1)/2}\state{A_L\ldots A_1},
\]
where $k$ is the number of fermionic fields in the trace.
The factor $(-1)^L$ is related to the gauge group. 
For gauge group $\grU(N)$ both parities are possible.
To restrict to gauge groups $\grSO(N)$ and $\grSp(N)$ 
one should consider only states with positive parity%
\footnote{This is related to the appearance of the invariant
tensor $d^{abc}$ in $\grSU(N)$, which has negative parity. 
The structure constant $f^{abc}$ has positive parity.}. 

\paragraph{Adjoint.}
We work with a complex algebra and 
will not require an adjoint operation here. 
Nevertheless, a short comment is in order.
For a real algebra it might seem difficult 
to construct self-adjoint generators.
With a meaningful adjoint operation
it is, however, straightforward and we will sketch how 
this should look like.
The scalar product of two states
$\state{A_1\ldots A_L}$, 
$\state{B_1\ldots B_L}$ should vanish
unless both are related by a cyclic permutation.
For two equal states the scalar product should not 
simply be $\pm 1$ 
(the sign is due to statistics), 
but it should be related to the conjugation class. 
When a state is written as $\state{(A_1\ldots A_{L/n})^n}$
with $n$ as large as possible, the
square norm should be $\pm n$.

\paragraph{Spin Chain.}
Alternatively, single-trace local operators can be viewed
as states of a cyclic super spin chain. The spin at each site 
can take five different alignments, where three are
to be considered `bosonic' and two are `fermionic'.
Note that the number of sites $L$ is not fixed for this 
\emph{dynamic} spin chain. The full Hilbert space is the 
tensor product of all Hilbert spaces of a fixed length $L$.

%%%%%%%%%%%%%%%%%%%%%%%%%%%%%%%%%%%%%%%%%%%%%%%%%%%%%%%%%%%%%%%%%%%%%%%%%%%%%
\subsection{The Algebra}

The fields $W_A$ transform canonically in a fundamental
$\rep{3|2}$ representation of $\alSU(2|3)$. 
Let us start by defining this algebra. 
The algebra consists of the generators 
(we enhance the symmetry algebra by one $\alU(1)$ generator)
\[
J=\{L^\alpha{}_\beta,R^a{}_b,H,\delta H;Q^a{}_\alpha,S^\alpha{}_a\}.
\]
The semicolon separates bosonic from fermionic operators.
The $\alSU(2)$ and $\alSU(3)$ generators $L^\alpha{}_\beta$ and $R^a{}_b$ 
are traceless, $L^\alpha{}_\alpha=R^a{}_a=0$.
The commutators are defined as follows.
Under the rotations $L,R$, the indices of any generator $J$
transform canonically according to the rules
\[\label{eq:su23rot}
\arraycolsep0pt
\begin{array}{rclcrcl}
\comm{L^\alpha{}_\beta}{J_\gamma}\eq
\delta^\alpha_\gamma J_\beta
-\half \delta^\alpha_\beta J_\gamma,
&\quad&
\comm{L^\alpha{}_\beta}{J^\gamma}\eq
-\delta^\gamma_\beta J^\alpha
+\half \delta^\alpha_\beta J^\gamma,
\\
\comm{R^a{}_b}{J_c}\eq
\delta^a_c J_b
-\sfrac{1}{3} \delta^a_b J_c,
&&
\comm{R^a{}_b}{J^c}\eq
-\delta_b^c J^a
+\sfrac{1}{3} \delta^a_b J^c.
\end{array}\]
The commutators of the full Hamiltonian $H$ and the interacting Hamiltonian 
$\delta H$ are given by
\[\label{eq:su23charge}
\comm{H}{J}=\eng(J)\, J,\qquad \comm{\delta H}{J}=0.
\]
This means that $\delta H$ is the central $\alU(1)$ generator and 
the non-vanishing energies are
\[\label{eq:su23dim}
\eng(Q)=-\eng(S)=\half.
\]
The supercharges anticommuting into rotations are given by%
\footnote{\label{fn:six}The linear combination $H+\half \delta H$ is 
a generator of $\alSU(2|3)$. Nevertheless, we would like 
to stick to these two generators, so that $H$ is
directly related to the dilatation generator $D$ in $\superN=4$ SYM,
see \secref{sec:N4toSU23}.}
\[\label{eq:su23momrot}
\acomm{S^\alpha{}_a}{Q^b{}_\beta}=
  \delta^b_a L^\alpha{}_\beta
  +\delta_\beta^\alpha R^b{}_a
  +\delta_a^b \delta_\beta^\alpha (\sfrac{1}{3}H+\sfrac{1}{6}\delta H).
\]
Furthermore, we demand a parity even algebra
\[\label{eq:su23parity}
PJP^{-1}=J, \quad\mbox{or}\quad \comm{P}{J}=0.
\]

It is straightforward to find the fundamental $\rep{3|2}$ 
representation acting on $W_A$ (we will do this explicitly in the next section). 
As states are constructed from the fundamental fields $W_A$ there is
an induced representation on the space of states; 
this is simply a tensor product representation and we will denote it by $J_0$.
The aim of this paper is to investigate
deformations $J(g)$ of the representation 
$J_0=J(0)$ around $g=0$. These deformations
are furnished in such a way that they are compatible
(i) with the $\alSU(2|3)$ algebra \eqref{eq:su23rot}-\eqref{eq:su23momrot} and 
(ii) with $\superN=4$ SYM field theory and Feynman diagrams. 

%%%%%%%%%%%%%%%%%%%%%%%%%%%%%%%%%%%%%%%%%%%%%%%%%%%%%%%%%%%%%%%%%%%%%%%%%%%%%%%%
\subsection{Representations}
\label{sec:reps}
In terms of representation theory, a state is characterised 
by the charges 
\[E+\half \delta E,\quad s,\quad [q,p].\]
where $E$ is the energy, 
$\delta E$ is the energy shift,
$s$ is \emph{twice} the $\alSU(2)$ spin
and
$[q,p]$ are the $\alSU(3)$ Dynkin labels.
These can be arranged into Dynkin labels of
$\alSU(2|3)$
\[w=[s;r;q,p],
\qquad r=\sfrac{1}{3}(E+\half \delta E)+\sfrac{1}{2}s-\sfrac{1}{3}p-\sfrac{2}{3}q,\]
however, instead of the fermionic label $r$ we prefer to give the
physically meaningful energy.
Representations are characterised by their highest weight.
The highest weight of the fundamental representation is
for instance
\[w\indup{F}=[0;0;0,1].\]
%

%%%%%%%%%%%%%%%%%%%%%%%%%%%%%%%%%%%%%%%%
\paragraph{Constituents.}
It is helpful to know how to construct a state with given 
(classical) charges $E_0,s,p,q$ and length $L$ from the 
fundamental fields $\phi_{1,2,3}$, $\psi_{1,2}$. 
The numbers of constituents of each kind are given by
\[\label{eq:Constit}
n_{\phi}=n_{1,2,3}=\left(\begin{array}{l}
L-\sfrac{2}{3}E_0+\sfrac{2}{3}p+\sfrac{1}{3}q\\
L-\sfrac{2}{3}E_0-\sfrac{1}{3}p+\sfrac{1}{3}q\\
L-\sfrac{2}{3}E_0-\sfrac{1}{3}p-\sfrac{2}{3}q
\end{array}\right),
\qquad
n_{\psi}=n_{4,5}=
\left(\begin{array}{l}
E_0-L+\half s\\
E_0-L-\half s
\end{array}\right).
\]
%

%%%%%%%%%%%%%%%%%%%%%%%%%%%%%%%%%%%%
\paragraph{Shortenings.}

A generic multiplet of $\alSU(2|3)$ has the dimension
\[\label{eq:MultDim}
(32|32)\times (s+1)\times \half (p+1)(q+1)(p+q+2).\]
For such a long (typical) multiplet the `unitarity' bound%
\footnote{We use the terminology of $\superN=4$ SYM even 
if some terms are inappropriate.}
applies
\[\label{eq:Bound}
E+\half \delta E>3+\sfrac{3}{2}s+p+2q.
\]
However, under certain conditions on $E$, 
the multiplet is shortened (atypical).
We find three conditions relevant to the spin chain.
The first one is the `half-BPS' condition% 
\footnote{In fact, $4$ out of $6$ supercharges annihilate the state.} 
\[\label{eq:CondHalf}
E+\half \delta E=p,\quad
n_{2}+n_{3}+n_{4}+n_{5}=0.
\]
%
%Such a multiplet has dimension $1+p(p+1)|p(p+1)$.
The second one is the `quarter-BPS' condition% 
\footnote{In fact, $2$ out of $6$ supercharges annihilate the state.
Multiplets of this kind have states belonging to the $\alSU(2)$ subsector
of just two complex bosonic fields $\phi_{1,2}$.}
\[\label{eq:CondQuarter}
E+\half \delta E=p+2q, \qquad
2n_{3}+n_{4}+n_{5}=0.
\]
Although a quarter-BPS multiplet is beyond the
unitarity bound, it can acquire an energy shift
if it joins with another multiplet to form a long one.
The last condition determines semi-short multiplets
\[\label{eq:CondSemi}
E+\half \delta E=3+\sfrac{3}{2}s+p+2q, \qquad
n_{3}+n_{5}=1.
\]
A long multiplet whose energy approaches the
unitarity bound \eqref{eq:Bound}
splits in two at \eqref{eq:CondSemi}.
If $s>0$, the highest weight of the 
upper semi-short submultiplet is shifted by 
\[\label{eq:SemiSemi}
E_0\mapsto E_0+\half,\quad
s\mapsto s-1,\quad
q\mapsto q+1,\quad
L\mapsto L+1.
\]
For $s=0$ the upper submultiplet is
quarter-BPS and its highest weight is shifted by%
%\footnote{In some respects a quarter BPS can be interpreted,
%alternatively, as a semi-short multiplet with spin $s=-1$ as 
%would be suggested by \eqref{eq:SemiSemi}.}
%
\[\label{eq:SemiBPS}
E_0\mapsto E_0+1,\quad
q\mapsto q+2,\quad
L\mapsto L+1.
\]
Multiplet shortening will turn out to be important later on.
This is because the generators that relate both submultiplets
must act as $\order{g}$ so that the multiplet can indeed split at $g=0$.

%%%%%%%%%%%%%%%%%%%%%%%%%%%%%%%%%%%%%%%%%%%
\paragraph{Fluctuations in Length.}

Note that all three bosons together have a vanishing 
$\alSU(3)$ charge and energy $3$. 
Similarly, both fermions have vanishing $\alSU(2)$ spin
and energy $3$, i.e.~the same quantum numbers
\[\label{eq:EqualCharges}
\phi_{[1}\phi_2\phi_{3]}\sim \psi_{[1}\psi_{2]}.
\]
Therefore one can expect there to
be fluctuations between these two configurations. 
In field theory these are closely related to 
the Konishi anomaly \cite{Konishi:1984hf,Konishi:1985tu}.
A state composed from $n_1\geq n_2\geq n_3$ bosons and
$n_4\geq n_5$ fermions can mix
with states
\[
(n_1-k,n_2-k,n_3-k;n_4+k,n_5+k),\qquad   -n_5\leq k\leq n_3.
\]
Note that the length $L=n_1+n_2+n_3+n_4+n_5$ decreases by $k$. 
We will refer to this aspect of the spin chain as \emph{dynamic}.

Length fluctuations are also interesting for multiplet shortenings.
The highest weight of a half-BPS or quarter-BPS multiplet 
has fixed length due to $n_3=n_5=0$.
For semi-short multiplets we have $n_3+n_5=1$. 
This means that there will be length fluctuations for the 
highest weight state. 
Two of the six supercharges transform a $\phi_3$ into a $\psi_{1,2}$.
Naively, both cannot act at the same time 
because there is only one $\phi_3$
(we will always have $n_3=1$),
and the multiplet becomes semi-short. 
However, we could simultaneously replace the resulting $\psi_{[1}\psi_{2]}$ 
by $\phi_{[1}\phi_2\phi_{3]}$ and thus fill up the $\phi_3$-hole.
A suitable rule is%
\footnote{In fact, this is part of the `classical' supersymmetry
variation (before a rescaling of fields by $g$).}
\[
Q^3{}_{2}\,\psi_1 \sim g\,\phi_{[1}\phi_{2]}.
\]
This is also the step between the two semi-short submultiplets
\eqref{eq:SemiSemi}. This property was recently used in 
\cite{Eden:2003sj} to determine two-loop 
scaling dimensions for operators at the unitarity bounds
from a one-loop field-theory calculation.
Note that $Q^3{}_{2}$ kills the highest weight 
when $n_4=0$; we need to apply $Q^3{}_{1}$ first to produce a $\psi_1$. 
In this case the upper submultiplet is
quarter-BPS \eqref{eq:SemiBPS}.
Furthermore note that when we apply $Q^3{}_1$ first, there are no 
more $\phi_3$'s and $\psi_2$'s left and length
fluctuations are ruled out. 
Therefore in a
BPS or semi-short multiplet we can always find 
states with fixed length.
In contrast, all states in a multiplet away 
from the unitarity bound \eqref{eq:Bound} are mixtures 
of states of different lengths.

%%%%%%%%%%%%%%%%%%%%%%%%%%%%%%%%%%%%%%%%%%%%%%%%%%%%%%%%%%%%%%%%%%%%%%%%%%%%%%%%
\subsection{From $\superN=4$ SYM to $\alSU(2|3)$}
\label{sec:N4toSU23}
A state of free $\superN=4$ SYM is characterised by the 
classical dimension $\oldDelta_0$, the $\alSU(2)^2$ labels
$[s,s_2]$, the $\alSU(4)$ Dynkin labels $[q,p,q_2]$,
the $\alU(1)$ hypercharge $B$ as well as the length $L$.
The $\alSU(2|3)$ subsector is obtained by restricting to states with 
\[\label{eq:EighthBPSCharges}
\oldDelta_0=p+\sfrac{1}{2}q+\sfrac{3}{2}q_2.
\]
This also implies $s_2=0$ and $\oldDelta_0=B+L$.
We write these as relations of the corresponding generators
\[\label{eq:SU2}
R^4{}_4=\half D_0,\quad 
\dot L^{\dot \alpha}{}_{\dot\beta}=0,\quad
D_0=L+B.
\]
Furthermore, we express the $\alSU(4)$ generator 
$R_{\alSU(4)}$ in terms of an $\alSU(3)$ generator $R$
\[
(R_{\alSU(4)})^a{}_b=R^a{}_b-\sfrac{1}{6}\delta^a_b D_0.
\]
Now we can reduce the $\alPSU(2,2|4)$ algebra 
as given in \cite{Beisert:2003jj} to the 
$\alSU(2|3)$ subsector and find 
precisely the $\alSU(2|3)$ relations 
\eqref{eq:su23rot}-\eqref{eq:su23momrot}
if the Hamiltonian $H$ is identified with 
the dilatation generator as follows
\[
H=D,\qquad \delta H=D-D_0.
\]
As we would like to directly compare to $\superN=4$ SYM
we write one of the generators of $\alSU(2|3)$ 
as $H+\half \delta H$ instead of
assigning a new letter.

We note that the states in this subsector are (classically) eighth-BPS in
terms of ${\superN=4}$ SYM. Unprotected primary states of the subsector
can therefore not be primary states of $\alPSU(2,2|4)$. 
The corresponding superconformal primaries have modified charges
\[
\oldDelta_0\mapsto E_0-1,\quad
q_2\mapsto\sfrac{2}{3}(\oldDelta_0-2-p-\half q),\quad
L\mapsto L-1.
\]
%

%%%%%%%%%%%%%%%%%%%%%%%%%%%%%%%%%%%%%%%%%%%%%%%%%%%%%%%%%%%%%%%%%%%%%%%%%%%%%%%%
\section{One-Loop}
\label{sec:OneLoop}

In this section we construct deformations of the algebra generators
$J(g)$ obeying the algebra relations \eqref{eq:su23rot}-\eqref{eq:su23momrot}. 
Here, we will proceed up to $\order{g^3}$
for the deformations of the Hamiltonian $H(g)$. 
This can still be done conveniently by hand 
without the help of computer algebra systems.
This section is meant to illustrate the methods of this paper
in a simple context before we proceed to 
higher loops in the sections to follow.

%%%%%%%%%%%%%%%%%%%%%%%%%%%%%%%%%%%%%%%%%%%%%%%%%%%%%%%%%%%%%%%%%%%%%%%%%%%%%%%%
\subsection{Interactions}

We write the interacting generators as a perturbation series in 
the coupling constant $g$
\[\label{eq:Jexpand}
J(g)=\sum_{k=0}^\infty g^k\, J_k.\]
The planar interactions within $J_k$ will be represented by the symbol
\[\label{eq:NotInt}
\PTerm{A_1\ldots A_n}{B_1\ldots B_m},\]
it replaces every occurrence of the sequence of fields $A_1\ldots A_n$
within a state by $B_1\ldots B_m$.
The sign is determined by the following ordering of fields and variations
\[
\PTerm{A_1\ldots A_n}{B_1\ldots B_m} \sim
W_{B_1}\ldots W_{B_n}\frac{\delta}{\delta W_{A_m}}\ldots \frac{\delta}{\delta W_{A_1}}.
\]
More explicitly, the action on a state $\state{C_1\ldots C_L}$ is
(via \eqref{eq:TraceCycl} we first shift the sequence to be replaced 
to the front)
\[\label{eq:NotAct}
\sum_{i=0}^{L-1} (-1)^{(C_1\ldots C_i)(C_{i+1}\ldots C_L)}\delta^{A_1}_{C_{i+1}}\ldots \delta^{A_n}_{C_{i+n}}
\state{B_1 \ldots B_m C_{i+n+1}\ldots C_L C_1 \ldots C_i}.
\]
A sample action is (we shift the trace back to its original position)
\[
\PTerm{\alpha b c}{c \alpha b} \state{142334452}=\state{134234452}+\state{242334415}.
\]
Throughout the paper, Latin (Greek) letters represent any of 
the bosonic (fermionic) fields $1,2,3$ ($4,5$).
The interaction $\PTerm{\alpha b c}{c \alpha b}$ therefore
looks for one fermion followed by two bosons within the trace.
Wherever they can be found these three fields are
permuted such that the last boson comes first,
next the fermion and the other boson last.

%%%%%%%%%%%%%%%%%%%%%%%%%%%%%%%%%%%%%%%%%%%
\paragraph{Feynman diagrams.}
The corrections $J_k$ to the algebra generators
have a specific form due to their origin in
Feynman diagrams. A connected diagram at $k$ orders of
the coupling constant has at most $k+2$ external legs;
$l$ internal index loops reduce the number of legs by $2l$, 
i.e.~$J_k\sim \PTerm{A_1\ldots A_n}{B_1\ldots B_m}$ with $n+m=k+2-2l$.
Due to gauge invariance of cyclic states, we can add 
a pair of legs to either side of the interaction%
\footnote{We assume, in general, that the interaction is shorter than the state. 
Special care has to be taken for short states, see \secref{sec:Short}.}
\[\label{eq:Gauge}
\PTerm{a A_1\ldots A_n}{a B_1\ldots B_m}
+(-1)^{(A_1\ldots A_n)(B_1\ldots B_m)}\PTerm{\alpha A_1\ldots A_n}{\alpha B_1\ldots B_m}
\mathrel{\hat{=}}\PTerm{A_1\ldots A_n}{B_1\ldots B_m}
\mathrel{\hat{=}}
\PTerm{A_1\ldots A_n a}{B_1\ldots B_m a}
+\PTerm{A_1\ldots A_n\alpha}{B_1\ldots B_m\alpha}.
\]
The additional legs act as spectators, they do not change the interaction.
For definiteness, this gauge invariance is used to increase the number of legs 
to its maximum value $k+2$ at $\order{g^k}$
\[\label{eq:Jlegs}
J_k\sim \PTerm{A_1\ldots A_n}{B_1\ldots B_m}\quad \mbox{with }n+m=k+2.
\]

\paragraph{Parity.}
The parity operation for interactions 
corresponding to \eqref{eq:TraceParity}
is
($i$ and $j$ are the numbers of fermions in $A_1\ldots A_n$ and
$B_1\ldots B_m$, respectively)
\[
P\PTerm{A_1\ldots A_n}{B_1\ldots B_m}P^{-1}=
(-1)^{n+m+i(i+1)/2+j(j+1)/2} \PTerm{A_n\ldots A_1}{B_m\ldots B_i}.
\]

\paragraph{Adjoint.}
An adjoint operation compatible with the adjoint operation on states
described above should interchange the two rows in the interaction symbol
\[
\PTerm{A_1\ldots A_n}{B_1\ldots B_m}^\dagger\sim 
\PTerm{B_1\ldots B_m}{A_1\ldots A_n}.
\]
%

%%%%%%%%%%%%%%%%%%%%%%%%%%%%%%%%%%%%%%%%%%%%%%%%%%%%%%%%%%%%%%%%%%%%%%%%%%%%%%%%
\subsection{Tree-level}

Let us illustrate the procedure for the generators at tree-level.
At tree-level, composite states transform in tensor product representations
of the fundamental representation $\rep{3|2}$.
The generators therefore act on one field at a time. 
We write down the most general form of generators
that respects $\alSU(3)\times\alSU(2)$ symmetry
\<\label{eq:TreeStruct}
R^{a}{}_{b}\eq c_1\PTerm{a}{b}+c_2\delta^a_b\PTerm{c}{c},
\nln
L^{\alpha}{}_{\beta}\eq c_3\PTerm{\alpha}{\beta}+c_4\delta^\alpha_\beta\PTerm{\gamma}{\gamma},
\nln
H_0\eq
 c_5\PTerm{a}{a}+c_6\PTerm{\alpha}{\alpha},
\nln
(Q_0)^{a}{}_{\alpha}\eq c_7\PTerm{a}{\alpha},
\nln
(S_0)^{\alpha}{}_{a}\eq c_8\PTerm{\alpha}{a}.
\>
The algebra relations have two solutions. One is the trivial 
solution $c_k=0$ corresponding to the trivial representation.
The other solution requires
\[\label{eq:TreeCoeff}
c_1=c_3=c_5=1,\quad
c_2=-\sfrac{1}{3},\quad
c_4=-\sfrac{1}{2},\quad
c_6=\sfrac{3}{2},\quad
c_7=e^{i\beta_1},\quad
c_8=e^{-i\beta_1}.
\]
As expected, we find that the bosons and fermions have classical energy
$1$ and $\sfrac{3}{2}$, respectively
\[
H_0=\PTerm{a}{a}+\sfrac{3}{2}\PTerm{\alpha}{\alpha}.
\]
The appearance of a free parameter $\beta_1$ is related to
a possible rescaling of the bosons and fermions.
This can be represented in terms of a similarity 
transformation on the generators 
\[\label{eq:TreeAlpha}
J_0\mapsto \exp\bigbrk{2i\beta_1 H_0}\,J_0\,
\exp\bigbrk{-2i\beta_1 H_0}.
\]
Obviously, the algebra relations \eqref{eq:Comm} are invariant under
such a transformation.
The only other $\alSU(3)\times\alSU(2)$ invariant similarity transformation
besides \eqref{eq:TreeAlpha} is
\[\label{eq:TreeEps}
J_0\mapsto \exp\bigbrk{i\beta_2 L}\,J_0\,
\exp\bigbrk{-i\beta_2 L},
\]
where $L$ is the length operator
\[\label{eq:Length}
L=\PTerm{a}{a}+\PTerm{\alpha}{\alpha}.
\]
The transformation \eqref{eq:TreeEps} is trivial 
and does not give rise to a new parameter 
at tree-level
because the length is conserved there
\[\label{eq:TreeLength}
\comm{L}{J_0}=0.
\]
%

%%%%%%%%%%%%%%%%%%%%%%%%%%%%%%%%%%%%%%%%%%%%%%%%%%%%%%%%%%%%%%%%%%%%%%%%%%%%%%%%
\subsection{Prediagonalisation}
Our aim is to diagonalise the full Hamiltonian $H(g)$. 
We cannot expect this to be possible on a basis of generators. 
However, we can diagonalise $H(g)$ 
\footnote{This bare Hamiltonian 
(of the BMN matrix model or of $\superN=4$ SYM on $\Real\times S^3$)
is the Legendre transform of the Lagrangian function and has the virtue
of terminating at $\gym^2$. Unfortunately, it does not 
conserve the classical dimension making it impossible to take 
the planar limit at the level of the Hamiltonian.
There would be diagrams with crossing lines which 
nevertheless would have to be considered as planar: 
A generator might, e.g.,~remove 
a set of fields without replacing them by anything else,
i.e.~$\PTerm{A_1\ldots A_n}{\cdot}$.
This could, \emph{a posteriori}, render a non-local insertion 
of some interaction planar by removing all 
fields in between.%
%Another indication is the nature of the
%'t~Hooft coupling $\lambda=\gym^2 N$, which implies
%that for every second interaction one index loop should be closed.
%For a single interaction we would have to close half an index loop
%which does not seem to make sense.
}
with respect to the 
classical Hamiltonian $H_0$.
This is because all elementary interactions have a definite energy
\[\label{eq:DimInter}
\bigcomm{H_0}{\PTerm{A_1\ldots A_n}{B_1\ldots B_m}}=
\eng\bigbrk{\PTerm{A_1\ldots A_n}{B_1\ldots B_m}} \,
\PTerm{A_1\ldots A_n}{B_1\ldots B_m},
\]
where the energy of the interaction is given by the
total energy of fields $B_1\ldots B_m$ minus the 
one of $A_1\ldots A_n$.
As in \cite{Kim:2002if,Kim:2003rz}
we can use a similarity transformation 
$H(g)\mapsto T(g)H(g)T^{-1}(g)$
to diagonalise $H(g)$ perturbatively with respect to $H_0$ such that%
\footnote{For that we can use the formula \eqref{eq:PerturbDiag},
although there is no clear notion of a an energy $e:=E_0$ of states 
when working with interaction symbols only.
The important point is that we
can keep track of the energy offset $E_0-H_0$ ($h:=H_0$)
using \eqref{eq:DimInter} when the interaction $V(g):=H(g)-H_0=\order{g}$ is applied.
This energy offset is the only essential quantity in the
propagator $\Delta_{E_0}$ \eqref{eq:PerturbProp}
and the projector $\Pi_{E_0}$ (which projects to zero energy offset).}
\[\label{eq:Ddiag}
\comm{H_0}{H(g)}=0.
\]
From now on we will assume that we have obtained \eqref{eq:Ddiag}.
This is not only convenient for reducing the number
of independent terms, but also necessary to have a well-defined 
notion of planar diagrams in terms of interactions.
Conservation of classical energy by $H(g)$ also
implies that the other interacting generators 
have definite classical energy%
\footnote{In \cite{Bianchi:2003wx,Beisert:2003te} it is shown
how to match the spectra of 
free $\superN=4$ SYM and tensionless
strings on $AdS_5\times S^5$.
The emergence of the operator $H_0$, 
which can be used to classify states according to
their classical energy $E_0$,
therefore should leave some traces also for
nearly tensionless strings.}
\[\label{eq:Jdiag}
\comm{H_0}{J(g)}=\eng(J) \,J(g),
\]
which can be shown as follows:
Let $\Pi_{E_0}$ project to the states of classical energy $E_0$.
Then $\Pi_{E_0}$ commutes with $H_m$ for arbitrary $E_0,m$
due to \eqref{eq:Ddiag}.
Now we project the algebra relation $\comm{H(g)}{J(g)}=\eng(J)\, J(g)$ to subspaces
of energy $E_0,E'_0$ and pick the $\order{g^n}$ contribution
\[\label{eq:JclassInd}
\sum_{m=1}^{n} \Pi_{E_0} \comm{H_{m}}{J_{n-m}} \Pi_{E_0'}
=\bigbrk{\eng(J)-(E_0-E'_0)}\Pi_{E_0} J_{n} \Pi_{E_0'}.
\]
We assume that $\comm{H_0}{J_m}=\eng(J)J_m$ for all $m<n$.
This is equivalent to the statement $\Pi_{E_0}J_m \Pi_{E'_0}=0$ for all $E_0-E'_0\neq\eng (J)$.
Choosing $E_0-E'_0\neq\eng (J)$ in \eqref{eq:JclassInd} we find that 
$\Pi_{E_0}J_n \Pi_{E'_0}$ must also vanish. The claim is by proved by induction.

We can now combine \eqref{eq:Jdiag} with the algebra relation
\eqref{eq:su23charge} and infer that the energy shift is 
conserved by the interacting algebra
\[\label{eq:AnoComm}
\comm{J(g)}{\delta H(g)}=0.
\]
%

%%%%%%%%%%%%%%%%%%%%%%%%%%%%%%%%%%%%%%%%%%%%%%%%%%%%%%%%%%%%%%%%%%%%%%%%%%%%%%%%
\subsection{First Order}
Let us restrict \eqref{eq:AnoComm} to its leading order
\[\label{eq:Leading}
\comm{J_0}{H_k}=0,
\]
in other words, the leading correction to the Hamiltonian
at some $\order{g^k}$ is conserved by the classical algebra.
We will now exclude a correction to $H$ at first order $k=1$
by representation theory: Due to \eqref{eq:Jlegs} the possible 
interactions are of the form 
\[\label{eq:NoFirst}
\PTerm{\cdot}{ABC},\quad\PTerm{C}{AB},\quad\PTerm{BC}{A},\quad\PTerm{ABC}{\cdot}.
\]
The indices cannot be fully contracted, hence there is no 
invariant interaction at $\order{g}$. In other words there
is no common irreducible representation of the free algebra
among the in and out channel
\[
\rep{1}\not\in(\rep{3|2})^3,\quad
\rep{3|2}\not\in(\rep{3|2})^2.
\]

%%%%%%%%%%%%%%%%%%%%%%%%%%%%%%%%%%%%%%%%%%%%%%%%%%%%%%%%%%%%%%%%%%%%%%%%%%%%%%%%
\subsection{Second Order}

\begin{figure}\centering
\parbox[c]{1.5cm}{\centering\includegraphics{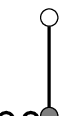}}
\parbox[c]{1.5cm}{\centering\includegraphics{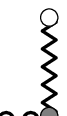}}
\parbox[c]{1.5cm}{\centering\includegraphics{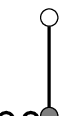}}
\parbox[c]{1.5cm}{\centering\includegraphics{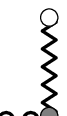}}
\parbox[c]{1.5cm}{\centering\includegraphics{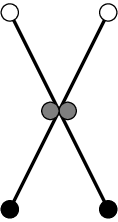}}
\parbox[c]{1.5cm}{\centering\includegraphics{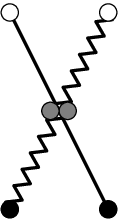}}
\parbox[c]{1.5cm}{\centering\includegraphics{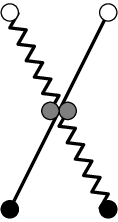}}
\parbox[c]{1.5cm}{\centering\includegraphics{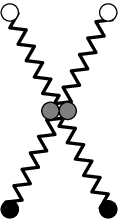}}
\caption{The structures for the construction of $H_2$. Straight and zigzag lines 
correspond to bosons and fermions, respectively.}
\label{fig:D2SU2SU3}
\end{figure}

A similar argument is used to show that at second order 
we must evenly distribute the four fields among the in and out channel,
i.e.
\[
\rep{1}\not\in(\rep{3|2})^4,\quad
\rep{3|2}\not\in(\rep{3|2})^3,\quad\mbox{but }
(\rep{3|2})^2=(\rep{3|2})^2.
\]
The most general form of $H_2$,
expressed as an action on bosons ($a,b$) and fermions ($\alpha,\beta$)
is therefore
\<\label{eq:D2}
H_2\eq
 c_1\PTerm{ab}{ab}
+c_2\PTerm{a\beta}{a\beta}
+c_2'\PTerm{\alpha b}{\alpha b}
+c_3\PTerm{\alpha \beta}{\alpha\beta}
\nl
+c_4\PTerm{ab}{ba}
+c_5\PTerm{a\beta}{\beta a}
+c_5'\PTerm{\alpha b}{b \alpha}
+c_6\PTerm{\alpha \beta}{\beta\alpha},
\>
see also \figref{fig:D2SU2SU3}.
First of all we demand that $H_2$ conserves parity \eqref{eq:su23parity},
\[
PH_2P^{-1}=H_2.
\]
As can be easily seen this requires
\[\label{eq:D2parity}
c_2=c_2',\quad c_5=c_5'.\]
We now commute $Q_0$ with $H_2$ and find 
\<\label{eq:DQ2}
\comm{(Q_0)^{a}{}_{\alpha}}{H_2}\eq
e^{i\beta_1}(c_1-c_2)\bigbrk{\PTerm{a b}{\alpha b}+\PTerm{ba}{b\alpha}}
+e^{i\beta_1}(c_4-c_5)\bigbrk{\PTerm{b a}{\alpha b}+\PTerm{ab}{b \alpha}}
\nlnum\nonumber
+e^{i\beta_1}(c_2-c_3)\bigbrk{\PTerm{a \beta}{\alpha \beta}-\PTerm{\beta a}{\beta\alpha}}
-e^{i\beta_1}(c_5+c_6)\bigbrk{\PTerm{a \beta}{\beta\alpha}-\PTerm{\beta a}{\alpha \beta}}.
\>
According to \eqref{eq:Leading} this must vanish, so we set
\[\label{eq:Coeff2}
c_1=c_2=c_3,\quad c_4=c_5=-c_6.
\]
The commutator $\comm{S_0}{H_2}$ leads to the same set of constraints.
\begin{figure}\centering
$H_2=c_1\parbox[c]{1.5cm}{\centering\includegraphics{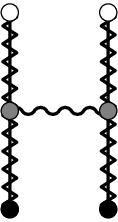}}+c_4
\parbox[c]{1.5cm}{\centering\includegraphics{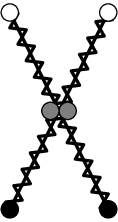}}$
\caption{The structures of $H_2$ which are compatible
with $\alSU(2|3)$ symmetry at leading order.
A straight+zigzag line correspond to a supermultiplet.}
\label{fig:D2SU23}
\end{figure}%
The two independent constants correspond to the two
irreducible representations in the tensor product,
see \figref{fig:D2SU23}
\[\label{eq:VFxVF}
\rep{3|2}\times \rep{3|2}=
\rep{7|6}\indup{+}+\rep{6|6}\indup{-}.
\]
More explicitly, $c_1+c_4$ corresponds to the symmetric product
and $c_1-c_4$ to the antisymmetric one.

%%%%%%%%%%%%%%%%%%%%%%%%%%%%%%%%%%%%%%%%%%%%%%%%%%%%%%%%%%%%%%%%%%%%%%%%%%%%%%%%
\subsection{Third Order}

The virtue of a classically invariant interaction 
applies only to the leading order, for $H_3$ we should break it.
However, we do not wish to break classical $\alSU(2|3)$ in the most general way,
but assume that the classical $\alSU(3)\times \alSU(2)$ invariance is conserved.
In field theory these would correspond to symmetries compatible with the 
regularisation scheme.

\begin{figure}\centering
$H_3=c_7\parbox[c]{2.5cm}{\centering\includegraphics{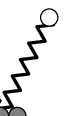}}
+c_8\parbox[c]{2.5cm}{\centering\includegraphics{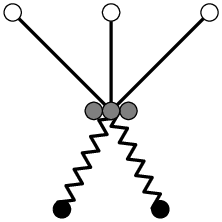}}$,\quad
$Q_1=c_9\parbox[c]{1.5cm}{\centering\includegraphics{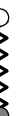}}$,\quad
$S_1=c_{10}\parbox[c]{1.5cm}{\centering\includegraphics{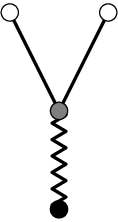}}$
\caption{The structures for the construction of $H_3,Q_1,S_1$.
The number of spin sites is not conserved here.}
\label{fig:D3}
\end{figure}
\begin{figure}
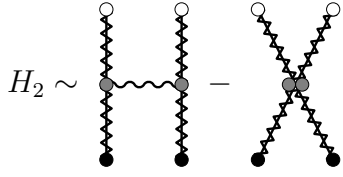
\centering
$H_2\sim
\parbox[c]{1.5cm}{\centering\includegraphics{su23.Vertex.2x.eps}}-
\parbox[c]{1.5cm}{\centering\includegraphics{su23.Vertex.2y.eps}}$
\caption{Closure of the algebra at $\order{g^3}$ fixes 
the relative coefficients within $H_2$.}
\label{fig:D2SU23final}
\end{figure}

The possible third order corrections involve the 
totally antisymmetric tensors of $\alSU(3)$ and $\alSU(2)$,
see \figref{fig:D3}:
\<
H_3\eq
c_7 \,\varepsilon_{\alpha\beta}\varepsilon^{abc}\PTerm{\alpha\beta}{abc}
+c_8 \,\varepsilon_{abc}\varepsilon^{\alpha\beta}\PTerm{abc}{\alpha\beta},
\nln
(Q_1)^{a}{}_{\alpha}\eq c_{9}\,\varepsilon_{\alpha\beta}\varepsilon^{abc}\PTerm{\beta}{bc},
\nln
(S_1)^{\alpha}{}_{a}\eq c_{10}\,\varepsilon_{abc}\varepsilon^{\alpha\beta}\PTerm{bc}{\beta}.
\>
With these expressions it is possible, yet tedious, to work out 
the commutators at third order by hand.
It is useful to note a version of the gauge invariance identity \eqref{eq:Gauge} adapted to 
this particular situation
\[
\PTerm{\beta d}{bcd}+\PTerm{\beta\delta}{bc\delta}=
\PTerm{d \beta}{dbc}-\PTerm{\delta\beta}{\delta bc}=
\PTerm{\beta}{bc}.
\]
Furthermore we will need some identities of the totally antisymmetric tensors 
$\epsilon^{abc}$ and $\epsilon_{\alpha\beta}$
and find for the commutator $\comm{Q}{\delta H}$ at $\order{g^3}$
\<\comm{(Q_0)^{a}{}_{\alpha}}{H_3}+\comm{(Q_1)^{a}{}_{\alpha}}{H_2}\eq
(c_4c_{9}-e^{i\beta_1}c_7)\,\varepsilon^{bcd}\varepsilon_{\alpha\beta}
\bigbrk{\PTerm{a\beta}{bcd}-\PTerm{\beta a}{bcd}}
\nl
+(c_4c_{9}-e^{i\beta_1}c_7)\,\varepsilon^{abc}\varepsilon_{\beta\gamma}
\bigbrk{-\PTerm{\beta \gamma}{\alpha bc}+\PTerm{\beta \gamma}{b \alpha c}-\PTerm{\beta \gamma}{bc \alpha}}
\nl
-(c_1+c_4)c_{9}\,\varepsilon^{abc}\varepsilon_{\alpha\beta}
\PTerm{\beta}{bc}.
\>
To satisfy \eqref{eq:AnoComm} this must vanish. 
The commutator $\comm{S}{\delta H}$ gives similar constraints and 
closure of the algebra requires
\[
c_1=-c_4,\quad e^{i\beta_1}c_7=c_4c_{9},\quad e^{-i\beta_1}c_8=c_4c_{10}.
\]
Here, there are two types of constraints. 
The latter two fix the 
coefficients of $Q_3$ and $S_3$. The first one
is more interesting, it fixes a coefficient of $H_2$ from one order below,
see \figref{fig:D2SU23final}.
This is related to the fact that 
$H_2$ was constructed to assign equal energy shifts
to all states of a multiplet of the free algebra.
In a superalgebra several atypical multiplets of the free theory
can join to form one typical multiplet in the interacting theory,
see \secref{sec:reps}.
A consistency requirement for this to happen is that the 
energy shift of the submultiplets agree.
In this case, it is achieved by $c_1=-c_4$.
To ensure agreement of energy shifts in terms of commutators,
we need to consider one additional power of the coupling constant, 
which is required to move between the submultiplets.
Furthermore, we note that the constraint $c_1=-c_4$ 
assigns a zero eigenvalue to the representation $\rep{7|6}$ in \eqref{eq:VFxVF}. 
This is essential, because $\rep{7|6}$ is in fact BPS and should not
receive an energy shift. 

%%%%%%%%%%%%%%%%%%%%%%%%%%%%%%%%%%%%%%%%%%%%%%%%%%%%%%%%%%%%%%%%%%%%%%%%%%%%%%%%
\subsection{Conclusions}

We now set the remaining independent constants $c_1,c_9,c_{10}$
to
\[ 
c_1=\alpha_1^2, \quad c_9=\sfrac{1}{\sqrt{2}}\, \alpha_1\, e^{i\beta_1+i\beta_2},\quad 
c_{10}=\sfrac{1}{\sqrt{2}}\, \alpha_1\, e^{-i\beta_1-i\beta'_2}.\]
In total we find the deformations at third order
\<
H_2\eq
 \alpha_1^2\PTerm{ab}{ab}
+\alpha_1^2\bigbrk{\PTerm{a\beta}{a\beta}+\PTerm{\alpha b}{\alpha b}}
+\alpha_1^2\PTerm{\alpha \beta}{\alpha\beta}
\nl
-\alpha_1^2\PTerm{ab}{ba}
-\alpha_1^2\bigbrk{\PTerm{a\beta}{\beta a}+\PTerm{\alpha b}{b \alpha}}
-\alpha_1^2\PTerm{\alpha \beta}{\beta\alpha},
\nln
H_3\eq
-\sfrac{1}{\sqrt{2}}\,\alpha_1^3\, e^{i\beta_2}\,\varepsilon_{\alpha\beta}\varepsilon^{abc}\PTerm{\alpha\beta}{abc}
-\sfrac{1}{\sqrt{2}}\,\alpha_1^3\, e^{-i\beta'_2}\,\varepsilon_{abc}\varepsilon^{\alpha\beta}\PTerm{abc}{\alpha\beta},
\nln
(Q_1)^{a}{}_{\alpha}\eq \sfrac{1}{\sqrt{2}}\,\alpha_1\, e^{i\beta_1+i\beta_2}\,\varepsilon_{\alpha\beta}\varepsilon^{abc}\PTerm{\beta}{bc},
\nln
(S_1)^{\alpha}{}_{a}\eq \sfrac{1}{\sqrt{2}}\,\alpha_1\, e^{-i\beta_1-i\beta'_2}\,\varepsilon_{abc}\varepsilon^{\alpha\beta}\PTerm{bc}{\beta}.
\>
Let us discuss the free parameters. 
The parameter $\beta'_2$ will in fact be determined 
by a constraint from fourth order: $c_9c_{10}=\half c_1$ or
\[\beta'_2=\beta_2.\]
As shown in \eqref{eq:TreeAlpha},\eqref{eq:TreeEps}, $\beta_{1,2}$ correspond 
to a similarity transformation of the algebra
\[
J_0\mapsto \exp\bigbrk{2i\beta_1 H_0+i\beta_2 L}\,J_0\,
\exp\bigbrk{-2i\beta_1 H_0-i\beta_2 L}.
\]
The algebra relations \eqref{eq:Comm} are invariant under similarity 
transformations, so $\beta_{1,2}$ can take arbitrary values.
For convenience, we might fix a gauge and set $\beta_1=\beta_2=0$,
but we will not do that here. 
Last but not least, the parameter $\alpha_1$ corresponds to a rescaling 
of the coupling constant
\[
g\mapsto \alpha_1\, g.
\]
The algebra relations \eqref{eq:Comm} are also invariant under this redefinition.

In a real form of the algebra we get a few additional constraints.
There, the algebra should be self-adjoint
which imposes some reality constraint on $\alpha_1,\beta_{1,2}$. 
For real $\alSU(2|3)$ they would have to be real and $\alpha_1^2$ would be positive. This
ensures positive planar energy shifts
as required by the unitarity bound.

In conclusion we have found that the deformations of the generators
are uniquely fixed at one-loop. 
Note that $H_2$ is proportional to the 
outcome of the harmonic action of \cite{Beisert:2003jj}.
Here, it is understood that some parameters cannot
be fixed due to symmetries of the algebra relations.
In determining the coefficients we saw that $\comm{\delta H(g)}{J(g)}=0$ 
at order $\order{g^{2\ell}}$ makes the $\ell$-loop energy shift
agree within semi-short multiplets whereas $\order{g^{2\ell+1}}$ joins 
up semi-short multiplets into long multiplets.
We have also seen that, for some purposes, the corrections 
to the Hamiltonian $H_{k}$ should be treated 
on equal footing as the other generators $Q_{k-2},S_{k-2}$ shifted by two orders. 
This can be attributed to the fact that commutation with $H_0$ is trivial 
and $H_1$ vanishes. We will refer to the generators
$H_{k},Q_{k-2},S_{k-2}$ collectively as $k$-th order.
Note that the anticommutator of supercharges at
`second order' is just the classical one
\[\acomm{(S_0)^{\alpha}{}_{a}}{(Q_0)^{b}{}_{\beta}}
=\delta^b_a L^\alpha{}_\beta
+\delta^\alpha_\beta R^b{}_a
+\sfrac{1}{3} \delta^b_a \delta^\alpha_\beta H_0.\]
At third order it is trivially satisfied due to the 
flavours of incoming and outgoing fields
\[\acomm{(S_1)^{\alpha}{}_{a}}{(Q_0)^{b}{}_{\beta}}
+\acomm{(S_0)^{\alpha}{}_{a}}{(Q_1)^{b}{}_{\beta}}
=
\half \delta^b_a \delta^\alpha_\beta H_1
=0.\]
%

%%%%%%%%%%%%%%%%%%%%%%%%%%%%%%%%%%%%%%%%%%%%%%%%%%%%%%%%%%%%%%%%%%%%%%%%%%%%%%%%
\section{Two-Loop}
\label{sec:TwoLoop}

In this section we will discuss the restrictions from 
the algebra at two-loops, i.e.~up to fifth order.
The steps are straightforward, but involve very lengthy expressions. 
We have relied on the algebra system \texttt{Mathematica} 
to perform the necessary computations.

%%%%%%%%%%%%%%%%%%%%%%%%%%%%%%%%%%%%%%%%%%%%%%%%%%%%%%%%%%%%%%%%%%%%%%%%%%%%%%%%
\subsection{Structures}

At fourth order we need to determine $H_4,Q_2,S_2$. 
For $H_4$ the $\alSU(3)\times \alSU(2)$ invariant interactions
which preserve the classical energy also preserve the number
of fields, i.e.~three fields are mapped into three fields.
Similarly for $Q_2,S_2$ we need two fields going into two fields
\[H_4\sim \PTerm{A_1A_2A_3}{B_1B_2B_3},\quad 
Q_2,S_2\sim \PTerm{A_1A_2}{B_1B_2}.\]
It is an easy exercise to count the number of
structures in $H_4,Q_2,S_2$. For $H_4$ there $2^3=8$
ways to determine the statistics of $A_1A_2A_3$ 
and $3!=6$ ways to permute the fields 
(each $A$ must be contracted to one of the $B$'s).
In total there are $6\cdot 8=48$ structures for $H_4$ and $8$ for $Q_2,S_2$ each.
We now demand parity conservation. This restricts the number of 
independent structures to $28$ and $4$ for $H_4$ and $Q_2,S_2$, respectively.

At fourth order we need to determine $H_5,Q_3,S_3$. 
Like $H_3,Q_1,S_1$, all of these involve the totally antisymmetric tensors for
$\alSU(3),\alSU(2)$ and change the number of fields by one. 
Counting of independent structures is also straightforward,
we find $48$ for $H_5$ and $12$ for $Q_3,S_3$ each.
Parity conservation halves each of these numbers.

%%%%%%%%%%%%%%%%%%%%%%%%%%%%%%%%%%%%%%%%%%%%%%%%%%%%%%%%%%%%%%%%%%%%%%%%%%%%%%%%
\subsection{Coefficients}

Now we demand that 
energy shifts are conserved at
fifth order
\[
\comm{Q^a{}_\alpha}{\delta H}=\comm{S^\alpha{}_a}{\delta H}=\order{g^6}.
\]
This fixes the remaining coefficient at third order and 
many coefficients at fourth and fifth order.
The anticommutator of supercharges \eqref{eq:su23momrot}
\[
\acomm{S^\alpha{}_a}{Q^b{}_\beta}=
  \delta^b_a L^\alpha{}_\beta
  +\delta_\beta^\alpha R^b{}_a
  +\delta_a^b \delta_\beta^\alpha (\sfrac{1}{3}H+\sfrac{1}{6}\delta H)+\order{g^4}
\]
does not lead to additional constraints.
The resulting deformations of the generators up to fourth order are presented
in \tabref{tab:twoloop}.
In the remainder of this subsection we shall discuss the 
undetermined coefficients.
\begin{table}
\<
R^{a}{}_{b}\eq \PTerm{a}{b}-\sfrac{1}{3}\delta^a_b\PTerm{c}{c},
\nln
L^{\alpha}{}_{\beta}\eq \PTerm{\alpha}{\beta}-\sfrac{1}{2}\delta^\alpha_\beta\PTerm{\gamma}{\gamma},
\nln
H_0\eq
 \PTerm{a}{a}+\sfrac{3}{2} \PTerm{\alpha}{\alpha},
\nln
H_2\eq
 \alpha_1^2\PTerm{ab}{ab}
+\alpha_1^2\bigbrk{\PTerm{a\beta}{a\beta}+\PTerm{\alpha b}{\alpha b}}
+\alpha_1^2\PTerm{\alpha \beta}{\alpha\beta}
\nl
-\alpha_1^2\PTerm{ab}{ba}
-\alpha_1^2\bigbrk{\PTerm{a\beta}{\beta a}+\PTerm{\alpha b}{b \alpha}}
+\alpha_1^2\PTerm{\alpha \beta}{\beta\alpha},
\nln
H_3\eq
-\sfrac{1}{\sqrt{2}}\,\alpha_1^3\,e^{i\beta_2}\,\varepsilon_{\alpha\beta}\varepsilon^{abc}\PTerm{\alpha\beta}{abc}
-\sfrac{1}{\sqrt{2}}\,\alpha_1^3\,e^{-i\beta_2}\,\varepsilon_{abc}\varepsilon^{\alpha\beta}\PTerm{abc}{\alpha\beta},
\nln
H_4\eq
(-2\alpha_1^4+2\alpha_1\alpha_3)\PTerm{a b c}{a b c}
+(-\half \alpha_1^4+2\alpha_1\alpha_3+\delta_3)\PTerm{\alpha \beta \gamma}{\alpha \beta \gamma}
\nl
+(\sfrac{1}{2}\alpha_1^4+2\alpha_1\alpha_3+2\delta_2)\PTerm{a \beta c}{a \beta c}
+(-4\alpha_1^4 +2\alpha_1\alpha_3- 2\delta_2)\PTerm{\alpha b\gamma}{\alpha b\gamma}
\nl
+(-\sfrac{11}{4}\alpha_1^4+2\alpha_1\alpha_3-\delta_2)\bigbrk{\PTerm{a b \gamma}{a b \gamma}+\PTerm{\alpha b c}{\alpha b c}}
+(2\alpha_1\alpha_3+\delta_2)\bigbrk{\PTerm{a \beta \gamma}{a \beta \gamma}+\PTerm{\alpha \beta c}{\alpha \beta c}}
\nl
+(\sfrac{3}{2}\alpha_1^4-\alpha_1\alpha_3)\bigbrk{\PTerm{a b c}{b a c}+\PTerm{a b c}{a c b}}
+(\alpha_1^4-\alpha_1\alpha_3)\bigbrk{\PTerm{a b \gamma}{b a \gamma}+\PTerm{\alpha b c}{\alpha c b}}
\nl
+(\sfrac{5}{4}\alpha_1^4-\alpha_1\alpha_3+i\alpha_1^2\gamma_3+i\delta_1)\bigbrk{\PTerm{\alpha b c}{b \alpha c}+\PTerm{a b \gamma}{a \gamma b}}
\nl
+(\sfrac{5}{4}\alpha_1^4-\alpha_1\alpha_3-i\alpha_1^2\gamma_3-i\delta_1)\bigbrk{\PTerm{a \beta c}{\beta a c}+\PTerm{a \beta c}{a c \beta}}
\nl
+(\alpha_1^4-\alpha_1\alpha_3+i\delta_1)\bigbrk{\PTerm{\alpha b \gamma}{b \alpha \gamma}+\PTerm{\alpha b \gamma}{\alpha \gamma b}}
\nl
+(\alpha_1^4-\alpha_1\alpha_3-i\delta_1)\bigbrk{\PTerm{a \beta \gamma}{\beta a \gamma}+\PTerm{\alpha \beta c}{\alpha c \beta}}
\nl
+(-\sfrac{7}{4}\alpha_1^4+\alpha_1\alpha_3)\bigbrk{\PTerm{\alpha \beta c}{\beta \alpha c}+\PTerm{a \beta \gamma}{a \gamma \beta}}
+(-\sfrac{7}{4}\alpha_1^4+\alpha_1\alpha_3-\delta_3)\bigbrk{\PTerm{\alpha \beta \gamma}{\beta \alpha \gamma}+\PTerm{\alpha \beta \gamma}{\alpha \gamma \beta}}
\nl
-\half \alpha_1^4 \bigbrk{\PTerm{a b c}{c a b}+\PTerm{a b c}{b c a}}
+\delta_3\bigbrk{\PTerm{\alpha \beta\gamma}{\gamma \alpha \beta}+\PTerm{\alpha \beta\gamma}{\beta \gamma \alpha}}
\nl
+(-\sfrac{1}{4}\alpha_1^4+i\alpha_1^2\gamma_1)\bigbrk{\PTerm{\alpha b c}{c \alpha b}+\PTerm{a b\gamma}{b \gamma a}}
+(\sfrac{1}{4} \alpha_1^4+i\alpha_1^2\gamma_2)\bigbrk{\PTerm{a \beta \gamma}{\gamma a \beta}+\PTerm{\alpha \beta c}{\beta c \alpha}}
\nl
+(-\sfrac{1}{4}\alpha_1^4-i\alpha_1^2\gamma_1)\bigbrk{\PTerm{a \beta c}{c a \beta}+\PTerm{a \beta c}{\beta c a}}
+(\sfrac{1}{4} \alpha_1^4-i\alpha_1^2\gamma_2)\bigbrk{\PTerm{\alpha b\gamma}{\gamma \alpha b}+\PTerm{\alpha b\gamma}{b \gamma \alpha}}
\nl
-\half \alpha_1^4 \bigbrk{\PTerm{a b \gamma}{\gamma b a}+\PTerm{\alpha b c}{c b \alpha}}
+\half \alpha_1^4 \PTerm{\alpha b \gamma}{\gamma b \alpha}
-\delta_3 \PTerm{\alpha \beta \gamma}{\gamma \beta \alpha},
\nln
(Q_0)^{a}{}_{\alpha}\eq e^{i\beta_1}\PTerm{a}{\alpha},
\nln
(Q_1)^{a}{}_{\alpha}\eq \sfrac{1}{\sqrt{2}}\,\alpha_1\,e^{i\beta_1+i\beta_2}\varepsilon_{\alpha\beta}\varepsilon^{abc}\PTerm{\beta}{bc},
\nln
(Q_2)^{a}{}_{\alpha}\eq 
e^{i\beta_1}(\sfrac{1}{4}\alpha_1^2-i\half \gamma_3+i\half \gamma_4)\bigbrk{\PTerm{a b}{\alpha b}+\PTerm{b a}{b \alpha}}
+e^{i\beta_1}(\half i\gamma_3+\half i\gamma_4)\bigbrk{\PTerm{a\beta}{\alpha\beta}-\PTerm{\beta a}{\beta\alpha}}
\nl
+e^{i\beta_1}(-\sfrac{1}{4} \alpha_1^2-i\gamma_1)\bigbrk{\PTerm{a b}{b \alpha}+\PTerm{b a}{\alpha b}}
+e^{i\beta_1}(\sfrac{1}{4} \alpha_1^2+i\gamma_2)\bigbrk{\PTerm{a \beta}{\beta \alpha}-\PTerm{\beta a}{\alpha \beta}},
\nln
(S_0)^{\alpha}{}_{a}\eq e^{-i\beta_1}\PTerm{\alpha}{a},
\nln
(S_1)^{\alpha}{}_{a}\eq \sfrac{1}{\sqrt{2}}\,\alpha_1\, e^{-i\beta_1-i\beta_2} \varepsilon_{abc}\varepsilon^{\alpha\beta}\PTerm{bc}{\beta},
\nln
(S_2)^{\alpha}{}_{a}\eq 
e^{-i\beta_1}(\sfrac{1}{4} \alpha_1^2+\half i\gamma_3-\half i\gamma_4)\bigbrk{\PTerm{\alpha b}{a b}+\PTerm{b \alpha}{b a}}
+e^{-i\beta_1}(-\half i\gamma_3-\half i\gamma_4)\bigbrk{\PTerm{\alpha \beta}{a \beta}-\PTerm{\beta \alpha}{\beta a}}
\nl
+e^{-i\beta_1}(-\sfrac{1}{4}\alpha_1^2+i\gamma_1)\bigbrk{\PTerm{\alpha b}{b a}+\PTerm{b \alpha}{a b}}
+e^{-i\beta_1}(\sfrac{1}{4}\alpha_1^2-i\gamma_2)\bigbrk{-\PTerm{\alpha \beta}{\beta a}+\PTerm{\beta \alpha}{a \beta}}.
\nonumber
\>

$H_4$ acting on two sites: 
$\bigeval{H_4}_{L=2}=
\sfrac{3}{2}\alpha_1^4\bigbrk{\PTerm{\alpha\beta}{\alpha\beta}-\PTerm{\alpha\beta}{\beta\alpha}}
+2\alpha_1\alpha_3 H_2$.

\caption{Two-loop deformations of the generators}
\label{tab:twoloop}
\end{table}

%%%%%%%%%%%%%%%%%%%%%%%%%%%%%%%%%%%%%%%%%%%%%%%%%%%%%%%%%%%%%%%%%%%%%%%%%%%%%%%%
\paragraph{Similarity transformations.}

As before, we can use a similarity transformation to
modify the generators
\[
J(g)\mapsto T(g)\,J(g)\,T(g)^{-1}.
\]
Before, we have used a transformation
which is independent of the coupling constant, 
here we will use a transformation proportional to $g^2$.
For consistency with the algebra the transformation 
will have to be $\alSU(3)\times \alSU(2)$ invariant and preserve 
the energy as well as parity. 
Also, according to \eqref{eq:Jlegs}, it should involve four fields. 
These are exactly the requirements for the form of $H_2$,
the $6$ independent structures are given 
in \eqref{eq:D2},\eqref{eq:D2parity}.
Of these six, there are two special combinations.
One is equivalent to the length operator
\[
L=\PTerm{ab}{ab}+\PTerm{a\beta}{a\beta}+\PTerm{\alpha b}{\alpha b}+\PTerm{\alpha\beta}{\alpha\beta}
\]
up to gauge transformations \eqref{eq:Gauge}
and the other is $H_2$ itself. 
The similarity transformation amounts to adding commutators with $H_2,Q_0,S_0$
\[
H_4\mapsto H_4+\comm{T_2}{H_2},\quad
J_2\mapsto J_2+\comm{T_2}{J_0}.
\]
These commutators vanish for $L$ and $H_2$
\[
\comm{H_2}{J_0}=\comm{L}{J_0}=\comm{H_2}{H_2}=\comm{L}{H_2}=0.
\]
This means that conjugation with $H_2$ and $L$ will have no effect on
$H_4,Q_2,S_2$.
The remaining four structures do not commute with $H_2,Q_0,S_0$
and amount to the constants $\gamma_{1,2,3,4}$. 
Note that $\gamma_4$ is related to the structure $H_0$ 
and does not appear in $H_4$ because of $\comm{H_0}{H_2}=0$.

%%%%%%%%%%%%%%%%%%%%%%%%%%%%%%%%%%%%%%%%%%%%%%%%%%%%%%%%%%%%%%%%%%%%%%%%%%%%%%%%
\paragraph{Gauge transformations.}

The constants $\delta_{1,2}$ multiply a structure
which is the difference
of the left hand side and right hand side in 
\eqref{eq:Gauge}. On cyclic states
its action is equivalent to zero.

%%%%%%%%%%%%%%%%%%%%%%%%%%%%%%%%%%%%%%%%%%%%%%%%%%%%%%%%%%%%%%%%%%%%%%%%%%%%%%%%
\paragraph{An $\alSU(2)$ identity.}

The constant $\delta_3$ multiplies a structure
which is zero due to an $\alSU(2)$ identity.
We cannot antisymmetrise more than two
fundamental representations of $\alSU(2)$
\[
\PTerm{\alpha\beta\gamma}{[\alpha\beta\gamma]}=0.
\]

%%%%%%%%%%%%%%%%%%%%%%%%%%%%%%%%%%%%%%%%%%%%%%%%%%%%%%%%%%%%%%%%%%%%%%%%%%%%%%%%
\paragraph{Transformations of the coupling constant.}

Finally, we are allowed to perform 
a transformation of the coupling constant 
\[\label{eq:Redef}
J(g)\mapsto J(f(g)).
\]
If we use the function $f(g)=\alpha_1 g+\alpha_3 g^3$
we find that 
\[
H_4\mapsto \alpha_1^4 H_4+2\alpha_1\alpha_3 H_2.
\]
%

%%%%%%%%%%%%%%%%%%%%%%%%%%%%%%%%%%%%%%%%%%%%%%%%%%%%%%%%%%%%%%%%%%%%%%%%%%%%%%%%
\subsection{Short states}
\label{sec:Short}

The fourth order interactions $H_4$ act on three fields.
We should also determine its action on the states of length two
\[
\Op_{(ab)}=\state{ab}=\Tr \phi_a\phi_b,\quad
\Op_{a\beta}=\state{a\beta}=\Tr \phi_a\psi_b,\quad
\Op_1=\epsilon^{\alpha\beta}\state{\alpha\beta}=\epsilon^{\alpha\beta}\Tr \psi_\alpha\psi_\beta.
\]
Together these form the protected half-BPS multiplet $\rep{7|6}$.
It is therefore reassuring to see that $\Op_{(ab)}$ and $\Op_{a\beta}$
are annihilated by $H_2,H_3$;
just as well they should be annihilated by $H_4$.
For $\Op_1$ the situation is different: It is annihilated by 
$H_2$, but $H_3$ produces the operator
\[
\Op_2=\epsilon^{abc}\state{abc}=\epsilon^{abc}\Tr\phi_a\phi_b\phi_c.
\]
The action of $H(g)$ on these two operators up to fourth order is given by
\[
H(g)=\matr{cc}{3+\epsilon g^4 & -2\sqrt{2}\,e^{i\beta_2}\alpha_1^3 g^3\\ 
-9\sqrt{2}\,e^{i\beta_2}e^{-i\beta_2} \alpha_1^3 g^3
&3+6\alpha_1^2 g^2-18 \alpha_1^4 g^4+12\alpha_1\alpha_3 g^4},
\]
where we assumed that $H_4 \Op_1=\epsilon\Op_1$.
The eigenvalues of this matrix at fourth order are given by 
\[
E_1=3+\epsilon g^4-6\alpha_1^4 g^4,\quad
E_2=3+6\alpha_1^2g^2-12\alpha_1^4g^4+12\alpha_1\alpha_3g^4.
\]
Due to its BPS nature the energy of the diagonalised $\Op_1$ 
must be exactly integer, 
$E_1=3$, and we set
\[
\epsilon=6\alpha_1^4.
\]
The fourth order interaction for states of length two is thus
\[
\bigeval{H_4}_{L=2}=
\sfrac{3}{2}\alpha_1^4\bigbrk{\PTerm{\alpha\beta}{\alpha\beta}-\PTerm{\alpha\beta}{\beta\alpha}}+
2\alpha_1\alpha_3 H_2.
\]
Here we have added a piece proportional to $H_2$, which is zero, 
for consistency with a redefinition of the coupling constant.

\begin{figure}\centering
\parbox[c]{3.5cm}{\centering\includegraphics{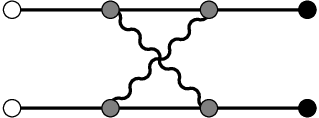}}
\parbox[c]{6.5cm}{\centering\includegraphics{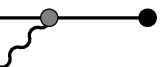}}
\caption{A generally non-planar interaction (left) can become
planar when acting on short states (right).}
\label{fig:nonplanar}
\end{figure}

At first sight, it may seem odd to have interactions which depend on the length of the
operator. From the point of view of Feynman diagrams this can actually happen: 
There are structures which are generically non-planar, 
but can act in a planar fashion on short states. 
In the notation of \cite{Beisert:2003tq} the relevant structure is,
\[
\Tr \bigcomm{\comm{\psi_\alpha}{T^m}}{\comm{T^n}{\check\psi^\alpha}}
\Tr \bigcomm{\comm{\psi_\beta}{T^n}}{\comm{T^m}{\check\psi^\beta}}.
\]
For example, this is a crossed two-gluon interaction of two scalar lines,
see \figref{fig:nonplanar}, left.
As the gluon lines are crossed, the interaction is non-planar. However, if
the interaction is connected to a state of two fields, one of the gluon lines
can be pulled around the state such that it does not cross any of the other lines,
see \figref{fig:nonplanar}, right. 
Therefore this interaction is planar exactly for states of length two.
This is quite interesting, because it shows that planar physics
does have some influence on (generally) non-planar physics.

%%%%%%%%%%%%%%%%%%%%%%%%%%%%%%%%%%%%%%%%%%%%%%%%%%%%%%%%%%%%%%%%%%%%%%%%%%%%%%%%
\subsection{Conclusions}

We see that for all free parameters in \tabref{tab:twoloop}
there is an associated symmetry of the algebra relations
and we can say that the two-loop contribution is uniquely fixed. 
The only parameter that influences energies is $\alpha_3$;
we cannot remove it by algebraic considerations.
The parameters $\gamma_{1,2,3,4}$ rotate only the eigenstates.
Finally, the parameters $\delta_{1,2,3}$ are there only because
we were not careful enough in finding \emph{independent} structures
(for $H_4$ there are only $25=28-3$ independent structures).
They have no effect at all. 

%%%%%%%%%%%%%%%%%%%%%%%%%%%%%%%%%%%%%%%%%%%%%%%%%%%%%%%%%%%%%%%%%%%%%%%%%%%%%%%%
%%%%%%%%%%%%%%%%%%%%%%%%%%%%%%%%%%%%%%%%%%%%%%%%%%%%%%%%%%%%%%%%%%%%%%%%%%%%%%%%
%%%%%%%%%%%%%%%%%%%%%%%%%%%%%%%%%%%%%%%%%%%%%%%%%%%%%%%%%%%%%%%%%%%%%%%%%%%%%%%%
\section{Three-Loop}
\label{sec:ThreeLoop}

For the sixth order contributions $H_6,Q_4,S_4$ we find in total 
$208+56+56$ parity conserving structures; they all conserve the number of fields%
\footnote{At eighth order the number of fields can be changed by two
using four epsilon symbols.}.
Of these only $173+32+32$ are independent due to identities
as discussed in the last section.
We imposed the constraint \eqref{eq:AnoComm} at sixth order 
\[\comm{Q}{\delta H}=\comm{S}{\delta H}=\order{g^7}.\]
and found that
the algebra relations fix $202$ coefficients (plus one
coefficient at fifth order). This leaves $35$ free coefficients.
The anticommutator of supercharges \eqref{eq:su23momrot}
at `sixth order'
\[
\acomm{S^\alpha{}_a}{Q^b{}_\beta}=
  \delta^b_a L^\alpha{}_\beta
  +\delta_\beta^\alpha R^b{}_a
  +\delta_a^b \delta_\beta^\alpha (\sfrac{1}{3}H+\sfrac{1}{6}\delta H)+\order{g^5}
\]
is satisfied automatically.

As we have learned above, the commutators at sixth order are not
enough to ensure consistency for splitting multiplets at the
unitarity bound, we should also consider seventh order. 
To preform those commutators would be even harder. 
We therefore consider a set of test multiplets at the unitarity bound. 
By requiring that the three-loop energy shifts coincide within 
submultiplets we are able to fix another $8$ coefficients. 

Still this leaves $27$ coefficients to be fixed, however, 
almost all of them rotate the space of states.
Experimentally, we found that only $4$ coefficients affect the energies. 
The remaining $23$ coefficients can be attributed to similarity transformations.
As before, the number of similarity transformations equals the number
of structures for $H_4$, i.e.~$25$. This means that there must be
$2$ commuting generators which are readily found to be $g^4 H_2$ and $g^4 L$. 
We summarise our findings concerning the number of 
coefficients in \tabref{tab:struct}. The symmetries indicated 
in the table refer to $L$ (which is conserved at second order but
broken at third order, hence the $-1$ at $k=3$), $g^2L$
(broken at fifth order), $\delta H$,
$g^4L$ (will break at seventh order) and $g^2 \delta H$.
This sequence of symmetries will continue at higher orders, but
there will be additional ones due to integrability
(see \secref{sec:Integ}):
Although the second integrable charge  
acts on three fields and commutes with $H_2$,
it does not appear here, because it is parity odd. 
The third charge is parity even and acts on at least four fields. 
Its commutator with $H_2$ vanishes and
it will therefore appear at eighth order as one additional symmetry.%
\begin{table}
\[\begin{array}{|c|rrrrr|}\hline
      k                           & 2 & 3 & 4 & 5 & 6   \\ \hline
H_{k}                             & 6 & 2 & 25& 18&173  \\ 
Q_{k-2}                           & 1 & 1 &  4&  6& 32  \\
S_{k-2}                           & 1 & 1 &  4&  6& 32  \\\hline
\mbox{total}                      & 8 & 4 & 33& 30&237  \\
\mbox{fixed at $\order{g^k}$}     & 5 & 2 & 25& 26&202  \\
\mbox{fixed at $\order{g^{k+1}}$} & 1 & 1 &  3&  1&  8  \\
\mbox{relevant}                   & 1 & 0 &  1&  0&  4  \\\hline
\mbox{irrelevant}                 & 1 & 1 &  4&  3& 23  \\
\mbox{symmetries}                 & 1 &-1 &  2& -1&  2  \\\hline
H_{k-2}                           & 2 & 0 &  6&  2& 25  \\
\hline
\end{array}\nonumber\]

\caption{Number of coefficients. $H_k,Q_{k-2},S_{k-2}$ give the number 
of independent structures that can be used for the construction of
generators.
The algebra relations fix a certain number of coefficients.
Of the remaining coefficients, some are relevant for energies
and some correspond to similarity transformations generated by 
the structures in $H_{k-2}$.
Some of the similarity transformations are symmetries.}
\label{tab:struct}
\end{table}

Let us now discuss the relevant coefficients.
One coefficient is due to a redefinition of the coupling constant and
cannot be fixed algebraically. 
To constrain the other three we will need further input.
Unfortunately the resulting generators are too lengthy to
be displayed here. Instead, let us have a look at the set 
of totally bosonic states.
In this subsector (which is closed only when further 
restricted to two flavours, i.e.~the $\alSU(2)$ subsector) 
$H_6$ is presented in \tabref{tab:H6su3}.

\begin{table}
\<
H_6\eq
\bigbrk{\sfrac{15}{2}\alpha_1^6+\sigma_1-2\sigma_2+12\sigma_3-2\sigma_4+\xi_2+\zeta_1}\PTerm{abcd}{abcd}\nl
+\bigbrk{-\sfrac{13}{4}\alpha_1^6+\alpha_1^3\alpha_3-\sfrac{1}{4}\sigma_1+\sigma_2-3\sigma_3+\sigma_4-\sfrac{1}{2}\xi_2-\zeta_1+\zeta_3}\bigbrk{\PTerm{abcd}{abdc}+\PTerm{abcd}{bacd}}\nl
+\bigbrk{-\sfrac{13}{2}\alpha_1^6+2\alpha_1^3\alpha_3-\sfrac{1}{2}\sigma_1+2\sigma_2-6\sigma_3+2\sigma_4-\zeta_1-\xi_2-2\zeta_3}\PTerm{abcd}{acbd}\nl
+\bigbrk{\sfrac{3}{2}\alpha_1^6-\alpha_1^3\alpha_3-\sigma_4+\sfrac{1}{2}\xi_2+\zeta_1-i\zeta_2}\bigbrk{\PTerm{abcd}{acdb}+\PTerm{abcd}{cabd}}\nl
+\bigbrk{\sfrac{3}{2}\alpha_1^6-\alpha_1^3\alpha_3-\sigma_4+\sfrac{1}{2}\xi_2+\zeta_1+i\zeta_2}\bigbrk{\PTerm{abcd}{adbc}+\PTerm{abcd}{bcad}}\nl
+\bigbrk{-\sigma_2-\sfrac{1}{2}\xi_2-\zeta_1}\bigbrk{\PTerm{abcd}{adcb}+\PTerm{abcd}{cbad}}
+\bigbrk{\sfrac{1}{2}\alpha_1^6-2\sigma_3+\zeta_1}\PTerm{abcd}{badc}\nl
+\bigbrk{-\sfrac{1}{2}\alpha_1^6+2\sigma_2-4\sigma_3+2\sigma_4-\zeta_1}\bigbrk{\PTerm{abcd}{bcda}+\PTerm{abcd}{dabc}}\nl
+\bigbrk{-2\sigma_2+4\sigma_3+\sigma_4-i\xi_3-\zeta_1}\PTerm{abcd}{bdac}
+\bigbrk{-2\sigma_2+4\sigma_3+\sigma_4+i\xi_3-\zeta_1}\PTerm{abcd}{cadb}\nl
+\bigbrk{\sigma_3-\sigma_4+i\xi_1+\zeta_1}\bigbrk{\PTerm{abcd}{bdca}+\PTerm{abcd}{dbac}}
+\bigbrk{\sigma_3-\sigma_4-i\xi_1+\zeta_1}\bigbrk{\PTerm{abcd}{cbda}+\PTerm{abcd}{dacb}}\nl
+\bigbrk{-2\sigma_3-2\sigma_4+\zeta_1}\PTerm{abcd}{cdab}
+\bigbrk{-\zeta_1}\bigbrk{\PTerm{abcd}{cdba}+\PTerm{abcd}{dcab}}\nl
+\bigbrk{2\sigma_4-\zeta_1}\PTerm{abcd}{dbca}
+\bigbrk{\zeta_1}\PTerm{abcd}{dcba}.
\nonumber
\>\vspace{-0.8cm}
\caption{$H_6$ in the bosonic subsector.}
\label{tab:H6su3}

\end{table}

The coefficients in \tabref{tab:H6su3} can be understood as follows.
The coefficients $\sigma_{1,2,3,4}$ are relevant.
One of them, $\sigma_1$, multiplies the structure $H_2$ 
(up to spectator legs \eqref{eq:Gauge})
and therefore corresponds to a redefinition of the
coupling constant 
as in \eqref{eq:Redef}.
The coefficients $\zeta_{1,2,3}$ multiply structures which are
actually zero, more explicitly, 
$\zeta_1$ multiplies $\PTerm{abcd}{[abcd]}$
and $\zeta_{2,3}$ can be gauged away using \eqref{eq:Gauge}.
Finally, the coefficients $\xi_{1,2,3}$ are related 
to similarity transformations and have no effect of the
energy shifts.

The crucial point is that we want $H_6$ to be generated by Feynman diagrams. 
Here we can make use of a special property of the scalar sector. 
The Feynman diagrams with the maximum number of eight legs do not 
have internal index loops.
In the planar case, such diagrams are iterated one-loop diagrams.
This means that we can only have three permutations of \emph{adjacent}
fields. The structures
\[
\PTerm{abcd}{cdab},\PTerm{abcd}{bdca},\PTerm{abcd}{dbac},
\PTerm{abcd}{cbda},\PTerm{abcd}{dacb},\quad
\PTerm{abcd}{cdba},\PTerm{abcd}{dcab},\PTerm{abcd}{dbca},
\quad\PTerm{abcd}{dcba}
\]
consist of four, five or six crossings of adjacent fields and are therefore excluded. 
We must set their coefficients to zero
\[\sigma_3=\sigma_4=0,\qquad \zeta_1=\xi_1=0.\]
The final relevant coefficient $\sigma_2$ multiplies a structure
$\charge_{3,0}$ which commutes with $H_2=\charge_{2,0}$.
At this point we cannot determine $\sigma_2$, but believe that 
it might be fixed due 
to the anticommutator \eqref{eq:su23momrot} $\acomm{Q}{S}$
at $\order{g^6}$ (four loops), see \secref{sec:Integ}.

%%%%%%%%%%%%%%%%%%%%%%%%%%%%%%%%%%%%%%%%%%%%%%%%%%%%%%%%%%%%%%%%%%%%%%%%%%%%%%%%
\section{The Spectrum and Integrability}
\label{sec:Spec}

In this section we fix the remaining degrees of freedom within $H$
and apply it to find the energy shifts of the first few
states in the spectrum. 

%%%%%%%%%%%%%%%%%%%%%%%%%%%%%%%%%%%%%%%%%%%%%%%%%%%%%%%%%%%%%%%%%%%%%%%%%%%%%%%%
\subsection{The Remaining Coefficients}
\label{sec:LastCoeffs}

First of all, we would like to fix the remaining relevant
coefficients. 
We cannot do this algebraically, because 
they correspond to symmetries of the 
commutation relations, namely a redefinition of the coupling
constant 
\[
g\mapsto f(g).
\]
Unlike the other symmetries, it has relevant consequences, 
it implies that energies are changed according to
\[
E(g)\mapsto E(f(g)).
\]
In order to match these degrees of freedom to 
$\superN=4$ SYM we should use some scaling dimension
that is known to all orders in perturbation theory. 
Namely in the planar BMN limit \cite{Berenstein:2002jq}
precise all-orders results for gauge theory have been obtained
\cite{Gross:2002su,Santambrogio:2002sb}.
In fact it is sufficient to use the scaling behaviour
for BMN states, their energies should depend only on 
\[
\lambda'=\frac{\lambda}{J^2}.
\]
If we redefine the coupling constant $g\sim \sqrt{\lambda}$
we obtain for $\lambda'$
\[
\lambda'\mapsto \frac{\tilde f(\lambda)}{J^2}
=\frac{\tilde f_1\lambda+\tilde f_2\lambda^2+\ldots}{J^2}
=\tilde f_1\lambda'+\tilde f_2 \lambda^{\prime\,2}J^2+\ldots\,\,.
\]
The problem is that all the higher expansion 
coefficients of $\tilde f$ yield divergent contributions in the BMN limit
$J\to\infty$
\[
\frac{\tilde f_k\lambda^k}{J^2}=\tilde f_k \lambda^{\prime\, k} J^{2k-2}\to \infty
\qquad\mbox{for }k\neq 1,
\]
and the only degree of freedom compatible with the BMN limit 
is to rescale $\lambda$ by a factor $\tilde f_1$.
In our model we must define the coupling constant $g\sim \sqrt{\lambda}$ in
such a way as to obtain good scaling behaviour of energies in the BMN limit.
This fixes $\alpha_3,\sigma_1$ and, gladly, also the single
coefficient at three-loops that could not be determined, $\sigma_2$
\[
\alpha_3=0,\quad \sigma_1=0,\quad \sigma_2=0.
\]
After that we can only rescale $\alpha_1$. This final degree of freedom is 
eliminated by knowing a single scaling dimension, e.g.~the one of the 
Konishi multiplet \cite{Anselmi:1997mq,Anselmi:1998ms,Bianchi:1999ge}
$\delta D=3\gym^2N/4\pi^4$,
or using the quantitative BMN energy formula \cite{Berenstein:2002jq}.
We now define
\[
g^2=\frac{\gym^2 N}{8\pi^2}
\]
this fixes $\alpha_1$ to unity
\[
\alpha_1=1.
\]

We conclude that \emph{the planar three-loop Hamiltonian is uniquely fixed} 
by the symmetry algebra, field theory and the BMN limit. 
Together with the fact that this model is a closed subsector of
$\superN=4$ SYM \cite{Beisert:2003jj} we have thus 
derived the dilatation generator in the $\alSU(2|3)$ subsector at three-loops.
Similarly, this model is a closed subsector of the BMN matrix model
and the two Hamiltonians must agree up to three loops
(after a redefinition of the coupling constant and provided that 
the BMN matrix model has a BMN limit).
Some comments:
\begin{list}{$\bullet$}{\itemsep0pt}
\item
We confirm the
three-loop results of \cite{Beisert:2003tq,Klose:2003qc}.
These are restricted to the $\alSU(2)$ subsector.
Here, we have extended the analysis to the $\alSU(2|3)$ subsector.

\item
The three-loop contribution in \cite{Beisert:2003tq} is based 
on the assumption of higher-loop integrability. 
Here, we relied only on established facts of gauge theory
and can thus actually \emph{derive} integrability, see \secref{sec:Integ}. 

\item
The symmetry algebra $\alSU(2|3)$ constrains the number of independent
coefficients considerably more than $\alSU(2)$. 
As opposed to $3$ (two-loop) and $6$ (three-loop) coefficients 
in the $\alSU(2)$ subsector
we find only $1$ and $2$ \footnote{possibly one of these two is
determined at four-loops, see \secref{sec:Integ}.} here.
It may be hoped for that the full symmetry algebra 
of $\superN=4$ SYM, $\alPSU(2,2|4)$, fixes 
the coefficients uniquely to all orders.

\item
Our result does not agree with
a recent result of near plane wave string theory \cite{Callan:2003xr}.
One will have to think how to make the near-BMN correspondence work
in this case.

\end{list}

%%%%%%%%%%%%%%%%%%%%%%%%%%%%%%%%%%%%%%%%%%%%%%%%%%%%%%%%%%%%%%%%%%%%%%%%%%%%%%%%
\subsection{Spectrum}

We are now ready to compute numerical values for 
some energy shifts. For this
we should consider the charges 
$E_0,s,p,q,L$ of a state 
and compute its constituents
according to \eqref{eq:Constit}.
These are arranged within a trace in all possible ways
\[\label{eq:AllOps}
\Op_n=
(
\Tr
\phi_1^{n_1}
\phi_2^{n_2}
\phi_3^{n_3}
\psi_1^{n_4}
\psi_2^{n_5},
\ldots
).
\]
Note that the length $L$ is not a good quantum number at
$\order{\gym^3}$, so we must include states of all admissible lengths
in \eqref{eq:AllOps}.
In practice this means that we may replace a complete set 
of bosons $\phi_1\phi_2\phi_3$ by a complete set of fermions
$\psi_1\psi_2$, \eqref{eq:EqualCharges}. Due to conservation of charges 
the Hamiltonian closes on this set of states
and we can evaluate its matrix elements% 
\footnote{Note that, although the 
Hilbert space is infinite-dimensional,
the Hamiltonian acts on a space of
fixed classical dimension $E_0$. 
Therefore the matrix $H_n{}^m(g)$ has a finite size.}
\[\label{eq:Hmatrix}
H(g)\, \Op_n=H_n{}^m(g)\, \Op_m.
\]
It is a straightforward task to find the eigenvalues and their perturbations. 
By construction the $\order{g^0}$ eigenvalue $E_0$ is universal for all
states $\Op_n$ and perturbation theory will start at $\order{g^2}$
\[\label{eq:PerturbSplit}
H(g)=E_0+g^2 \lrbrk{h+V(g)},\qquad
\mbox{with }V=\order{g}.
\]
Diagonalising the leading order matrix $h$ is a non-linear problem. 
The resulting eigenvalues represent the one-loop energy shifts $\delta E_2$. 
Now we pick an eigenvalue $e$ of $h$ and consider the subspace of states
with one-loop energy shift $\delta E_2=e$.
The higher-order energy shifts are given by
(in contrast to the results of \cite{Klose:2003qc} 
$V'$ was constructed such that conjugation symmetry is preserved)
\<\label{eq:PerturbDiag}
V'\eq 
\smash{\sum\nolimits_{e}\Pi_e\Big[}
 V 
+ V\Delta_e V 
+ \bigbrk{V \Delta_e V\Delta_e V
-\half V \Delta_e^2 V\Pi_e V
-\half V \Pi_e V\Delta^2_e V}
\nl\qquad\qquad
+V\Delta_e V\Delta_e V\Delta_e V
-\half V\Delta_e^2 V\Pi_e V\Delta_e V
-\half V\Delta_e V\Pi_e V\Delta_e^2 V
\nl\qquad\qquad 
-\half V\Delta_e^2 V\Delta_e V\Pi_e V
-\half V\Pi_e V\Delta_e  V\Delta_e^2 V
\nlnum\qquad\qquad
-\half V\Delta_e V\Delta_e^2 V\Pi_e V
-\half V\Pi_e V\Delta_e^2 V\Delta_e V
\nl\qquad\qquad
+\sfrac{1}{3} V\Delta_e^3 V\Pi_e V\Pi_e  V
+\sfrac{1}{3} V\Pi_e V\Delta_e^3 V\Pi_e  V
+\sfrac{1}{3} V\Pi_e V\Pi_e V\Delta_e^3 V
+\ldots \smash{\Big]} \Pi_e.
\nonumber
\>
The propagator $\Delta_e$ is given by
\[\label{eq:PerturbProp}
\Delta_e= \frac{1-\Pi_e}{e-h}
\]
and $\Pi_e$ projects to the subspace
with leading correction $e$.
If there is only a single state with one-loop energy shift $e$,
\eqref{eq:PerturbDiag} gives its higher order corrections. 
For degenerate states at one-loop, 
\eqref{eq:PerturbSplit},\eqref{eq:PerturbDiag} must be applied 
iteratively until the resulting matrix $V'$ becomes diagonal%
\footnote{In principle it could happen that states
with equal one-loop energy shift have matrix
elements at $\order{g^3}$. In this case the energy
shift would have an expansion in terms of $g\sim\sqrt{\lambda}$ 
instead of $g^2\sim \lambda$
similar to the peculiarities noticed in \cite{Beisert:2003tq}.
It would be interesting to see if this 
does indeed happen or, if not, why?}.

Next, it is important to know the multiplets
of states. In the interacting theory there
are two types of single-trace multiplets, half-BPS and
long ones.
The half-BPS multiplets are easily identified,
there is one multiplet with labels
\[
E=E_0=L=p,\quad [0;0;0,p],\quad P=(-1)^p
\]
for each $p$, they receive no corrections to their energy.
Long multiplets are not so easy to find. 
By means of a \texttt{C++} computer programme
we explicitly constructed the spectrum of 
all states (up to some energy bound)
and iteratively removed the multiplets
corresponding to the leftover highest weight state
(this `sieve' algorithm, also reminiscent of the
standard algorithm for division, is described in more
detail in \cite{Bianchi:2003wx,Beisert:2003te}).
For a set of states with given charges as in \eqref{eq:AllOps}
this also tells us how many representatives there are 
from each of the multiplets and allows us to
identify the energy shift we are interested in.

Finally, to obtain the energy shift of a given multiplet, 
a lot of work can be saved by choosing a suitable representative. 
Resolving the mixing problem for the highest weight state 
is usually more involved than for a descendant. 
For instance, highest weight states involve all
three flavours of bosons, $n_1,n_2,n_3\geq 1$.
This increases the number
of permutations in \eqref{eq:AllOps}
and also gives rise to mixing between states 
of different lengths. 
The matrix $H_n{}^m$ \eqref{eq:Hmatrix} will be unnecessarily large.
If, instead, one applies three supergenerators
$Q^1{}_4 Q^2{}_4 Q^3{}_4$, i.e.
\[
n_1\mapsto n_1-1,\quad
n_2\mapsto n_2-1,\quad
n_3\mapsto n_3-1,\quad
n_4\mapsto n_4+3,
\]
the state becomes more uniform.
This decreases the number of permutations and,
in the case of multiplets at the unitarity bound \eqref{eq:Bound},
mixing between states of different lengths is
prevented due to $n_3=0$.

\begin{table}\centering
$\begin{array}{|l|lc|l|}\hline
E_0&\multicolumn{1}{c}{\alSU(2|3)}&L&\bigbrk{\delta E_2,-\delta E_4,\delta E_6}^P\\\hline
2  &[0;0;0,2]^{\dagger\bullet}&2&(0,0,0)^+ \\
\hline
3  &[0;0;0,3]^{\dagger\bullet}&3&(0,0,0)^- \\
3  &[0;1;0,0]^{\ast\bullet}&3&\bigbrk{6,12,42}^+ \\
\hline
4  &[0;0;0,4]^{\dagger\bullet}&4&(0,0,0)^+ \\
4  &[0;1;0,1]^{\ast\bullet}&4&\bigbrk{4,6,17}^- \\
\hline
5  &[0;0;0,5]^{\dagger\bullet}&5&(0,0,0)^- \\
5  &[0;1;0,2]^{\ast\bullet}&5&\bigbrk{10\omega-20,17\omega-60,\frac{117}{2}\omega-230}^+ \\
5  &[0;1;1,0]^{\ast\bullet}&5&\bigbrk{6,9,\frac{63}{2}}^- \\
\hline
6  &[0;0;0,6]^{\dagger\bullet}&6&(0,0,0)^+ \\
6  &[0;1;0,3]^{\ast\bullet}&6&\bigbrk{2,\frac{3}{2},\frac{37}{16}}^-,\bigbrk{6,\frac{21}{2},\frac{555}{16}}^- \\
6  &[0;1;1,1]^{\ast\bullet}&6&\bigbrk{5,\frac{15}{2},25}^\pm \\
6  &[0;2;0,0]&6&\bigbrk{14\omega-36,24\omega-90,\frac{173}{2}\omega-315}^+\\
6  &[2;3;0,0]^\ast&5&\bigbrk{10,20,\frac{145}{2}}^-\\
\hline
6.5&[1;2;0,2]^\ast&6&\bigbrk{8,14,49}^\pm \\
\hline
7  &[0;0;0,7]^{\dagger\bullet}&7&\bigbrk{0,0,0}^- \\
7  &[0;1;0,4]^{\ast\bullet}&7&\bigbrk{\scriptstyle 14\omega^2-56\omega+56,\,23\omega^2-172\omega+224,\,79\omega^2-695\omega+966}^+ \\
7  &[0;1;1,2]^{\ast\bullet}&7&\bigbrk{4,5,14}^\pm,\bigbrk{6,9,33}^- \\
7  &[0;1;2,0]^{\ast\bullet}&7&\bigbrk{\scriptstyle 20\omega^2-116\omega+200,\,32\omega^2-340\omega+800,\,112\omega^2-1400\omega+3600}^+ \\
7  &[0;2;0,1]&7&\bigbrk{\scriptstyle22\omega^2-144\omega+248,\,37\omega^2-460\omega+1016,\,125\omega^2-1893\omega+4438}^- \\
7  &[2;3;0,1]^\ast&6&\bigbrk{8,14,46}^+ \\
\hline
7.5&[1;2;0,3]^\ast&7&\bigbrk{7,12,\frac{83}{2}}^\pm \\
7.5&[1;2;1,1]^\ast&7&\bigbrk{6,\frac{33}{4},\frac{1557}{64}}^\pm,\bigbrk{10,\frac{75}{4},\frac{4315}{64}}^\pm \\
7.5&[1;3;0,0]&7&\bigbrk{9,15,51}^\pm \\
\hline
8  &[0;0;0,8]^{\dagger\bullet}&8&\bigbrk{0,0,0}^+ \\
8  &[0;1;0,5]^{\ast\bullet}&8&\bigbrk{4,5,\frac{49}{4}}^-,
                              \bigbrk{8\omega-8,13\omega-18,\frac{179}{4}\omega-61}^- \\
8  &[0;1;1,3]^{\ast\bullet}&8&\bigbrk{\scriptstyle 17\omega^2-90\omega+147,\,\frac{51}{2}\omega^2-\frac{525}{2}\omega+\frac{1239}{2},\,\frac{169}{2}\omega^2-\frac{2091}{2}\omega+\frac{5649}{2}}^\pm \\
8  &[0;1;2,1]^{\ast\bullet}&8&\bigbrk{5,\frac{15}{2},\frac{55}{2}}^\pm,
                              \bigbrk{12\omega-24,18\omega-54,57\omega-171}^- \\
8  &[0;2;0,2]&8&\bigbrk{7,\frac{19}{2},\frac{59}{2}}^\pm,
                \bigbrk{\scriptstyle 44\omega^5-768\omega^4+6752\omega^3-31168\omega^2+70528\omega-60224,\,
                        A,\,B}^+ \\
8  &[0;2;1,0]&8&\bigbrk{9,\frac{31}{2},\frac{103}{2}}^\pm,
                \bigbrk{\scriptstyle 24\omega^2-172\omega+344,\,39\omega^2-524\omega+1372,\,138\omega^2-2209\omega+6198}^- \\
8  &[2;3;0,2]^\ast&7&\bigbrk{\scriptstyle28\omega^2-252\omega+728,\,51\omega^2-906\omega+3864,\,179\omega^2-3965\omega+20090}^- \\
8  &[2;3;1,0]^\ast&7&\bigbrk{8,\frac{25}{2},\frac{687}{16}}^+,\bigbrk{12,\frac{45}{2},\frac{1281}{16}}^+ \\
\hline
8.5&[1;2;0,4]^\ast&8&\bigbrk{6,\frac{19}{2},\frac{247}{8}}^\pm,\bigbrk{8,\frac{29}{2},\frac{427}{8}}^\pm \\
8.5&[1;2;1,2]^\ast&8&\bigbrk{\scriptstyle 31\omega^3-350\omega^2+1704\omega-3016,\,
                             50\omega^3-1111\omega^2+7971\omega-18452,\,
                             C}^\pm \\
8.5&[1;2;2,0]^\ast&8&\bigbrk{8,13,\frac{343}{8}}^\pm,
                     \bigbrk{15\omega-48,23\omega-135,\frac{595}{8}\omega-\frac{4023}{8}}^\pm \\
8.5&[1;3;0,1]&8&\bigbrk{8,13,\frac{173}{4}}^\pm,
                \bigbrk{10,\frac{67}{4},\frac{3725}{64}}^\pm,
                \bigbrk{\scriptstyle 19\omega-86,\,\frac{133}{4}\omega-\frac{1169}{4},\,\frac{7395}{64}\omega-\frac{79503}{64}}^\pm \\
\hline
\end{array}$
{\scriptsize\<
A\eq 73\omega^5-2486\omega^4+31804\omega^3-188280\omega^2+506048\omega-487104\omega\nln
B\eq 251\omega^5-10452\omega^4+156202\omega^3-1041992\omega^2+3055168\omega-3125328\nln
C\eq \textstyle\frac{337}{2}\omega^3-\frac{18363}{4}\omega^2+38740\omega-102390\nonumber
\>\vspace{-1cm}}%
\caption{Spectrum of highest weight states with $E_0\leq 8.5$ in the 
dynamic $\alSU(2|3)$ spin chain. }
\label{tab:anotab}
\end{table}

We summarise our findings for states of classical 
energy $E_0\leq 8.5$ in \tabref{tab:anotab}.
We have labelled the states by their classical energy $E_0$, 
$\alSU(2|3)$ Dynkin labels, and classical length $L$.
For each multiplet we have given its energy shifts
$\delta E=g^2\delta E_2+g^4\delta E_4+g^6\delta E_6+\order{g^8}$
up to three-loops
(note that the two-loop energy shift is always negative,
we present its absolute value) and parity $P$. 
A pair of degenerate states with opposite parity
is labelled by $P=\pm$. 
For convenience we have indicated the shortening conditions
relevant for the $\alSU(2|3)$ representations:
Half-BPS multiplets and multiplets at the unitarity bound (which split at $g=0$) 
are labelled by $^\dagger$ and $^\ast$, respectively. 
For $s=0$ some of their components are in the $\alSU(2)$ subsector. 
Such multiplets are indicated
by $^\bullet$ and their energy shifts agree with the results
of \cite{Beisert:2003tq}.

Generically, the one-loop energy shifts are not fractional numbers
but solutions to some algebraic equations. We refrain from solving
them (numerically), but instead give the equations. 
In the table such states are indicated as polynomials 
$\delta E_{2,4,6}(\omega)$ of degree $k-1$ 
(note that the table shows $-\delta E_4(\omega)$). 
The energy shifts are obtained as solutions to the equation
\[
\delta E=g^2\, \omega,\qquad \omega^k=\delta E_2(\omega)+g^2\,\delta E_4(\omega)+g^4\,\delta E_6(\omega)+
\order{g^6}.
\]
%

%%%%%%%%%%%%%%%%%%%%%%%%%%%%%%%%%%%%%%%%%%%%%%%%%%%%%%%%%%%%%%%%%%%%%%%%%%%%%%%%
\subsection{Integrability}
\label{sec:Integ}

A closer look at \tabref{tab:anotab}
shows that there is a maximum amount of parity pairs.
All multiplets with opposite parities which can pair up,
actually do have degenerate energies. At the one-loop level,
where the Hamiltonian is the Hamiltonian of a classical 
integrable $\alSU(2|3)$ spin chain,
this is explained by integrability \cite{Beisert:2003tq}. For an integrable spin chain
one can construct a transfer operator $\transfer_0(u)$ satisfying 
\[
\comm{J_0}{\transfer_0(u)}=\comm{\transfer_0(u)}{\transfer_0(v)}=0.
\]
When expanded in the spectral parameter $u$ the transfer matrix 
\[
\transfer_0(u)=\exp \sum_{n=2}^\infty \frac{u^{n-1}}{(n-1)!}\,\charge_{n,0}
\]
gives rise to an infinite tower (for sufficiently long states) 
of mutually commuting local charges $\charge_{n,0}$,
\[
\comm{J_0}{\charge_{n,0}}=\comm{\charge_{m,0}}{\charge_{n,0}}=0.
\]
The charge $\charge_{n,0}$ extends over $n$ nearest neighbours
and has parity $(-1)^n$.
The one-loop Hamiltonian is the first non-trivial of these $H_2=\charge_{2,0}$. 
The second non-trivial charge, $\charge_{3,0}$, is parity odd and 
its action relates states of opposite parity
\[
\charge_{3,0}\,\Op_\pm \sim \Op_\mp.
\]
Its conservation implies the degeneracy of $\Op_\pm$.
In the classical spin chain this is only valid at one-loop, 
but here we notice that energy shifts 
do agree within pairs to the order we are investigating.
As shown in \cite{Beisert:2003tq} this can be explained 
by an extension of the higher charges $\charge_{n,0}$ to higher
loops 
\[\charge_n(g)=\sum_{k=0}^\infty g^k\,\charge_{n,k},\]
where $\charge_{n,k}$ is a local interaction with 
$2n+k$ legs (incoming plus outgoing fields).
Higher-loop integrability demands that the charges commute among themselves 
(in a perturbative sense)
\[\label{eq:IntAlg}
\comm{J(g)}{\charge_n(g)}=\comm{\charge_m(g)}{\charge_n(g)}=0
\]
or in terms of a higher-loop transfer operator
\[
\transfer(u,g)
=\sum_{k=0}^\infty g^{k}\,\transfer_{k}(u)
=\exp \sum_{n=2}^\infty \sum_{k=0}^\infty\frac{u^{n-1}}{(n-1)!}\,g^k\,\charge_{n,k}
\]
with 
\[
\comm{J(g)}{\transfer(u,g)}=\comm{\transfer(u,g)}{\transfer(v,g)}=0.
\]
Moreover, in analogy to \eqref{eq:Jdiag}, we should demand that the charges have zero classical energy
\[
\comm{H_0}{\charge_n(g)}=\comm{H_0}{\transfer(u,g)}=0.
\]
From \eqref{eq:IntAlg} it follows that the energy shift $\delta H(g)$ commutes with the charges 
\[\label{eq:deltaHinRinf}
\comm{\delta H(g)}{\charge_n(g)}=\comm{\delta H(g)}{\transfer(u,g)}=0.
\]

Here we did not construct any higher charges, but 
note that all possible pairs ($^\pm$) remain degenerate
at three-loops.
This is so for the pairs of 
the $\alSU(2)$ sector ($^\bullet$) \cite{Beisert:2003tq},
for pairs at the unitarity bound ($^\ast$), but
also, and most importantly, for pairs
away from the unitarity bound (unmarked in \tabref{tab:anotab}). 
As discussed at the end of \secref{sec:reps}
all states of such a multiplet are superpositions of 
states of different lengths.
We take this as compelling evidence for the 
existence of higher charges $\charge_n(g)$ 
\footnote{Experience with the $\alSU(2)$ subsector
shows that, as soon as all possible pairs
degenerate, we can not only 
construct one higher charge $\charge_3(g)$
to explain the degeneracies,
but also all the other commuting charges $\charge_n(g)$.
This parallels earlier observations that it 
appears close to impossible to construct systems with 
$\comm{\charge_2}{\charge_3}=0$ which are not integrable.}.
This is interesting, because it shows
that also for truly \emph{dynamic spin chains}
with a fluctuating number of sites 
integrability is an option.

Let us now comment on the r\^ole of the interaction Hamiltonian $\delta H(g)$. 
On the one hand it belongs to the symmetry algebra $\alSU(2|3)$ when combined 
with $H_0$ (see footnote \ref{fn:six} on page \pageref{fn:six})
\[\label{eq:HinSU23}
H(g)+\half \delta H(g)=H_0+\sfrac{3}{2}\delta H(g)\in \alSU(2|3).
\]
On the other hand \eqref{eq:deltaHinRinf}, together 
with $\comm{J(g)}{\delta H(g)}=0$ \eqref{eq:AnoComm}, 
shows that $\delta H(g)$ is also a generator of the 
abelian algebra of local integrable charges $\Real^\infty$
defined by \eqref{eq:IntAlg}
\[\label{eq:HinRinf}
g^{-2}\,\delta H(g)\in \Real^\infty.
\]

At this point we can reinvestigate the undetermined coefficients
at three-loops. By imposing
\[\label{eq:QSHthreeloop}
\comm{Q(g)}{\delta H(g)}=\comm{S(g)}{\delta H(g)}=\order{g^8}
\]
we found that $H_6$ depends on four relevant coefficients 
$\sigma_{1,2,3,4}$. The coefficients $\sigma_{3,4}$ multiply invalid structures, 
whereas $\sigma_1$ corresponds to a redefinition of the coupling constant. 
The remaining coefficient $\sigma_2$ multiplies 
$\charge_{4,0}$ which is structurally equivalent to $H_6$
and satisfies $\comm{J(g)}{\charge_4(g)}=0$ as well.
In fact, by imposing $\comm{Q(g)}{X(g)}=\comm{S(g)}{X(g)}=0$ we not only find $\delta H$, but 
also all the other even generators $\charge_{2n}(g)$ of the abelian algebra
of integrable charges $\Real^\infty$.
Thus $\sigma_2$ corresponds to the transformation
\[\label{eq:sigma2Q4}
\delta H(g)\mapsto \delta H(g)+\sigma_2 \,g^6\, \charge_{4}(g),
\]
which has no influence on \eqref{eq:QSHthreeloop} due to 
$\comm{J(g)}{\charge_n(g)}=0$. 
This degree of freedom may be fixed by considering the 
anticommutator of supercharges \eqref{eq:su23momrot} 
\[\label{eq:su23momrot7}
\acomm{S^\alpha{}_a(g)}{Q^b{}_\beta(g)}=
  \delta^b_a L^\alpha{}_\beta
  +\delta_\beta^\alpha R^b{}_a
  +\delta_a^b \delta_\beta^\alpha \bigbrk{\sfrac{1}{3}H_0+\sfrac{1}{2}\delta H(g)}
+\order{g^7}.
\]
Unfortunately this involves $Q_6,S_6$ which are part of a four-loop calculation and 
out of reach here. We believe that \eqref{eq:su23momrot7} will force the
coefficient $\sigma_2$ to vanish, and in conclusion
all corrections up to three loops would be determined
uniquely (up to a redefinition of the coupling constant).
At higher loops this picture is expected to continue:
While $\comm{Q(g)}{X(g)}=\comm{S(g)}{X(g)}=0$ determines
elements $\charge_{2n}$ of $\Real^\infty$, the anticommutator 
$\acomm{S(g)}{Q(g)}$ yields the one element $\delta H(g)$ \eqref{eq:HinRinf}
which is also associated to $\alSU(2|3)$ 
via \eqref{eq:HinSU23}.

Let us comment on the uniqueness of the $n$-th charge $\charge_n$.
As the charges form an abelian algebra, one can replace $\charge_n$ by 
some polynomial in the charges without changing the algebra.
This, however, would make $\charge_n$ multi-local in general. To preserve locality
we must use linear transformations which are generated by 
\[
\charge_n(g)\mapsto \charge_n(g)+\alpha\, g^{2k}\, \charge_{n+2m}(g).
\]
On the one hand, the structural constraint on $\charge_n$,
i.e.~that $\charge_n$ has at least as many legs as $g^{2k} \charge_{n+2m}$, 
requires $k\geq 2m$. On the other hand, it was observed that
$\charge_{n,k}$ scales as $\order{L^{-k-n}}$ 
in the thermodynamic/BMN limit $L\to\infty$ \cite{Arutyunov:2003rg} 
\[
\charge'_{n,k}=\lim_{L\to\infty} L^{k+n}\charge_{n,k}\quad\mbox{or}\quad
\transfer'(u',g')=\lim_{L\to\infty}\transfer(Lu',Lg')^L
\]
In order not to spoil this, we need $k\leq m$. Together, these two
constraints imply $k=m=0$, or, 
in other words, $\charge_k$ can only be rescaled by a constant.
Finally, this constant can be fixed by using the standard transfer 
matrix $\transfer_0(u)$ of the one-loop spin-chain, i.e.~there
is a canonical definition for $\charge_n(g)$.
As $g^{-2}\delta H(g)$ obeys the same constraints as $\charge_2(g)$,
both of them must be equal 
\[g^{-2}\,\delta H(g)=\charge_2(g).\]
The thermodynamic constraint does not allow the transformation 
\eqref{eq:sigma2Q4} and thus fixes $\sigma_2$ as shown in \secref{sec:LastCoeffs}
(even if it is expected to be determined at four-loops, see above).

%At one-loop this is manifest due to $H_2=\charge_{2,0}$, but note 
%that there is no reason to demand $g^{-2}\delta H(g)=\charge_2(g)$, 
%it might be more convenient to use a different linear combination
%%
%\[\label{eq:HasQk}
%g^{-2}\,\delta H(g)=\sum_{k=1}^\infty c_{2k}(g)\,\charge_{2k}(g),
%\]
%%
%which is ultimately determined by the anticommutator 
%of supercharges $\acomm{S}{Q}$, \eqref{eq:su23momrot}.
%In particular, there appears to be no need to fully `improve' the 
%semi-improved charges of \cite{Arutyunov:2003rg}
%if the expansion of $\mathcal{E}(\mu)$ in powers of $\mu$ is
%interpreted in a different way: The leading order of 
%$\mathcal{E}(\mu)$ might be interpreted as the energy $E$
%and the higher orders give the higher charges 
%$\charge_k$ starting with $\charge_2$
%(in \cite{Arutyunov:2003rg} the first higher charge
%was assumed to be $\charge_4$ demanding for 
%a further improvement).
%In this set of charges there is one redundancy due
%to \eqref{eq:HinSU23} and \eqref{eq:HinRinf} 
%(assuming $H_0$ to be trivial).
%On the one hand the redundancy would be removed by a full improvement.
%On the other hand the semi-improved result in \cite{Arutyunov:2003rg} 
%can be used to provide a `natural' definition of the $\charge_k$'s
%as well as the coefficients $c_k$ in \eqref{eq:HasQk}.

In agreement with the findings of Klose and Plefka
\cite{Klose:2003qc} we conclude that 
integrability appears to be a consequence
of field theory combined with symmetry and does not
depend on the specific model very much.
This strongly supports the idea
of all-loop integrability in planar 
$\superN=4$ SYM. What is more, the dynamic aspects
of higher-loop spin chains appear to 
be no obstruction.

%%%%%%%%%%%%%%%%%%%%%%%%%%%%%%%%%%%%%%%%%%%%%%%%%%%%%%%%%%%%%%%%%%%%%%%%%%%%%%%%%
%\section{The non-planar extension}
%\label{sec:NonPlanar}

%\remark{Draw colour diagrams,}
%\remark{reduce by Jacobi}

%\remark{Assign in/out,}
%\remark{reduce by Jacobi}

%\remark{Assign statistics}

%%%%%%%%%%%%%%%%%%%%%%%%%%%%%%%%%%%%%%%%%%%%%%%%%%%%%%%%%%%%%%%%%%%%%%%%%%%%%%%%%
%\section{Higher charges}
%\label{sec:Higher}

%%%%%%%%%%%%%%%%%%%%%%%%%%%%%%%%%%%%%%%%%%%%%%%%%%%%%%%%%%%%%%%%%%%%%%%%%%%%%%%%
\section{Outlook}
\label{sec:Outlook}

In this work we have derived the planar
dilatation generator of $\superN=4$ SYM in 
the $\alSU(2|3)$ subsector. This is 
equivalent to the Hamiltonian of a
\emph{dynamic} spin chain
where the number of spin sites is allowed to fluctuate.
There are a few open questions and some
directions for future research.

First of all, one could try to understand the two-loop
deformations in \tabref{tab:twoloop}
better and find out how they can be constructed
from scratch without having to start with the most general 
structure and subsequently constraining it.
Furthermore, it would be good to have 
some of the higher charges explicitly. 
Unfortunately, like the three-loop contribution,
already the second charge at two-loops is lengthy.
A consideration of the non-planar, finite $N$,
algebra would also be interesting. 

For the derivation of the three-loop result we have made use 
of the thermodynamic/BMN limit. 
As explained, this can probably be avoided by going to four-loops. 
Why does the closure of the algebra together with the
structural constraints yield a Hamiltonian $\delta H(g)$
with a good thermodynamic/BMN limit? What about the 
higher charges $\charge_n(g)$?

One might try to rederive the complete one-loop 
dilatation operator of $\superN=4$ gauge theory \cite{Beisert:2003jj} with these
algebraic methods. The closure of the corresponding 
$\alPSU(2,2|4)$ algebra at third order will probably 
reduce the infinite number of independent coefficients to just one.
One could then proceed to two-loops,
either in the full $\superN=4$ theory or 
in a subsector with non-compact representations.

It might also be interesting to extend the current analysis to
non-perturbative effects like instantons. 
Possibly the symmetry algebra also puts constraints 
on these and a direct computation as in \cite{Kovacs:2003rt}
might be simplified or even bypassed. 

In any case, a better understanding of 
dynamic spin chains and, even more
urgently, an improved notion of higher-loop integrability
is required. This might be a key to unravel 
planar $\superN=4$ gauge theory at all loops.

%%%%%%%%%%%%%%%%%%%%%%%%%%%%%%%%%%%%%%%%%%%%%%%%%%%%%%%%%%%%%%%%%%%%%%%%%%%%%%%%
\subsection*{Acknowledgements}

The author would like to thank 
Gleb Arutyunov, Volodya Kazakov, Thomas Klose, Jan Plefka,
Arkady Tseytlin
and, in particular, Matthias Staudacher 
for interesting discussions and useful comments.
Ich danke der \emph{Studienstiftung des
deutschen Volkes} f\"ur die Unterst\"utzung durch ein 
Promotions\-f\"orderungsstipendium.

\bibliography{su23}
\bibliographystyle{nb}

\end{document}